\documentclass[journal]{IEEEtran}
\IEEEoverridecommandlockouts
\usepackage{amsmath,amssymb,amsfonts}
\usepackage[table,xcdraw]{xcolor}
\usepackage{threeparttable}
\usepackage{svg}
\usepackage{xspace}
\usepackage{booktabs} 
\usepackage{tikz}
\usetikzlibrary{shapes, arrows, snakes}
\tikzset{>=latex}

\definecolor{ao}{rgb}{0.0, 0.5, 0.0}

\def\BibTeX{{\rm B\kern-.05em{\sc i\kern-.025em b}\kern-.08em
    T\kern-.1667em\lower.7ex\hbox{E}\kern-.125emX}}
    
\newcommand{\rebuttal}[1]{{#1}}

\usepackage{standalone}
\usepackage{anyfontsize}

\usepackage[utf8]{inputenc}
\usepackage{amsmath}
\usepackage{amsopn}
\usepackage{accents}
\usepackage{siunitx}
\usepackage{amssymb}
\usepackage{pifont}
\usepackage{mathrsfs} 
\usepackage{eucal} 
\renewcommand{\mathcal}[1]{\CMcal{#1}} 
\usepackage{trfsigns} 
\usepackage{nicefrac} 
\usepackage{url}

\usepackage{mleftright} 
\mleftright 

\newcommand{\Reals}{\mathbb{R}}      
\newcommand{\Binary}{\mathbb{B}}

\newcommand{\mimann}{\textsc{MI-mann}\xspace}

\DeclareUnicodeCharacter{2212}{-}

\usepackage{lineno}

\usepackage{fancyhdr}

\fancypagestyle{mahmood}{%
   \fancyhf{} 
   
   \fancyhead[C]{Accepted in IEEE Journal on Emerging and Selected Topics in Circuits and Systems (JETCAS)}
}%

\makeatletter
\let\ps@IEEEtitlepagestyle\ps@mahmood
\makeatother

\begin{document}


\title{Binarization Methods for Motor-Imagery Brain--Computer Interface Classification}




\author{Michael~Hersche, 
        Luca~Benini, 
        and~Abbas~Rahimi
\thanks{M. Hersche, and L. Benini are with the Integrated Systems Laboratory, ETH Zurich, 8092 Zurich, Switzerland (e-mail: hersche@iis.ee.ethz.ch).}
\thanks{L. Benini is also with the Department of Electrical, Electronic and Information Engineering, University of Bologna, 40136, Italy.}
\thanks{A. Rahimi is with IBM Research-Zurich, CH-8803 Zürich, Switzerland (e-mail: abr@zurich.ibm.com).}
\thanks{Manuscript received July XX, XXXX; revised September YY, YYYY.}
}

\maketitle

\begin{abstract}
Successful motor-imagery brain--computer interface (MI-BCI) algorithms either extract a large number of handcrafted features and train a classifier, or combine feature extraction and classification within deep convolutional neural networks (CNNs).
Both approaches typically result in a set of real-valued weights, that pose challenges when targeting real-time execution on tightly resource-constrained devices.
We propose methods for each of these approaches that allow transforming real-valued weights to binary numbers for efficient inference.
Our first method, based on sparse bipolar random projection, projects a large number of real-valued Riemannian covariance features to a binary space, where a linear SVM classifier can be learned with binary weights too.  
By tuning the dimension of the binary embedding, we achieve almost the same accuracy in 4-class MI ($\leq$1.27\% lower) compared to models with float16 weights, yet delivering a more compact model with simpler operations to execute.
Second, we propose to use memory-augmented neural networks (MANNs) for MI-BCI such that the augmented memory is binarized.
Our method replaces the fully connected layer of CNNs with a binary augmented memory using bipolar random projection, or learned projection.
Our experimental results on EEGNet, an already compact CNN for MI-BCI, show that it can be compressed by 1.28$\times$ at iso-accuracy using the random projection.
On the other hand, using the learned projection provides 3.89\% higher accuracy but increases the memory size by 28.10$\times$.
\end{abstract}
\begin{IEEEkeywords}
EEG, binary embedding, sparse random projection, SVM, binarized memory-augmented neural networks.
\end{IEEEkeywords}


\section{Introduction}\label{sec:intro}

\IEEEPARstart{B}{rain--computer} interfaces (BCIs) enable a communication channel between a user and an external device through intentional modulation of brain signals, e.g., motor imagery (MI) of movement of a part of the body~\cite{Ramadan2017BrainReview}.  
A BCI aims at recognizing human intentions from the analysis of spatiotemporal neural activity, typically recorded non-invasively by a number of electroencephalogram (EEG) electrodes.
Such information can enable controlling games~\cite{Saeedi2016AdaptiveReliability,Perdikis2018TheUsers}, driving a wheelchair~\cite{Xiong2019AActivation}, and even motor rehabilitation after stroke~\cite{Cho2019MotorMethod}.

Accurate EEG decoding of MI is a challenging task due to inter- and intra-subject variabilities~\cite{Tangermann2012ReviewIV.,Lotte2018AUpdate}.
Most approaches train a personalized model per subject to deal with the high variability of EEG signals between subjects~\cite{Schirrmeister2017DeepVisualization,Hersche2018,KaiKengAng2008FilterInterface,Barachant2013ClassificationApplications}.
Traditional approaches use well-known filter bank common spatial patterns (FBCSP)~\cite{KaiKengAng2008FilterInterface}, or Riemannian covariance features~\cite{Barachant2013ClassificationApplications} followed by an SVM or LDA classifier.
%
Among them, multi-spectral and temporal unsupervised Riemannian features with a linear SVM classifier~\cite{Hersche2018} achieve the highest average classification accuracy (75.47\%) among nine subjects on the 4-class BCI competition IV-2a dataset~\cite{Brunner2008BCIA}.
%
%

Recently, convolutional neural networks (CNNs) have gained increasing attention in the MI-BCI field, reducing the data pre-processing steps and eliminating the procedure of hand-crafting features.  
One of the first successful CNN in MI classification was FBCSP-inspired Shallow ConvNet~\cite{Schirrmeister2017DeepVisualization}.
The recent TPCT network~\cite{Li2020AElectrodes} achieves the state-of-the-art (SoA) accuracy of 88.87\% on the 4-class MI BCI Competition IV-2a dataset.
However, it requires a large number of 7.78\,M trainable weights and 1.73\,G multiply-accumulate (MAC) operations in inference.
In contrast, more compact models such as EEGNet~\cite{Lawhern2018EEGNet:Interfaces} provide a good trade-off between the number of trainable parameters, complexity, and accuracy. 

%

%
%

\begin{figure*}[htbp]
    \centering
    \footnotesize
    \includegraphics[width=\textwidth]{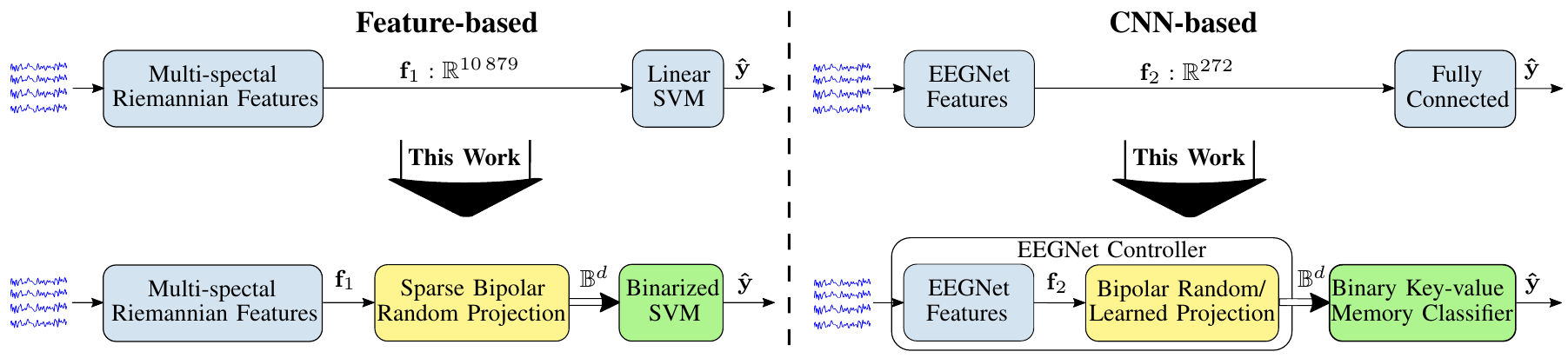}
    \caption{This work binarizes real-valued features in two common MI-classification approaches to d-dimensional Hamming space $\Binary^d$ with help of sparse/dense bipolar random projections or learned projections.}
    \label{fig:overview}
\end{figure*}   

\begin{figure}[htbp]
    \centering
    \footnotesize
    \includegraphics[width=\linewidth]{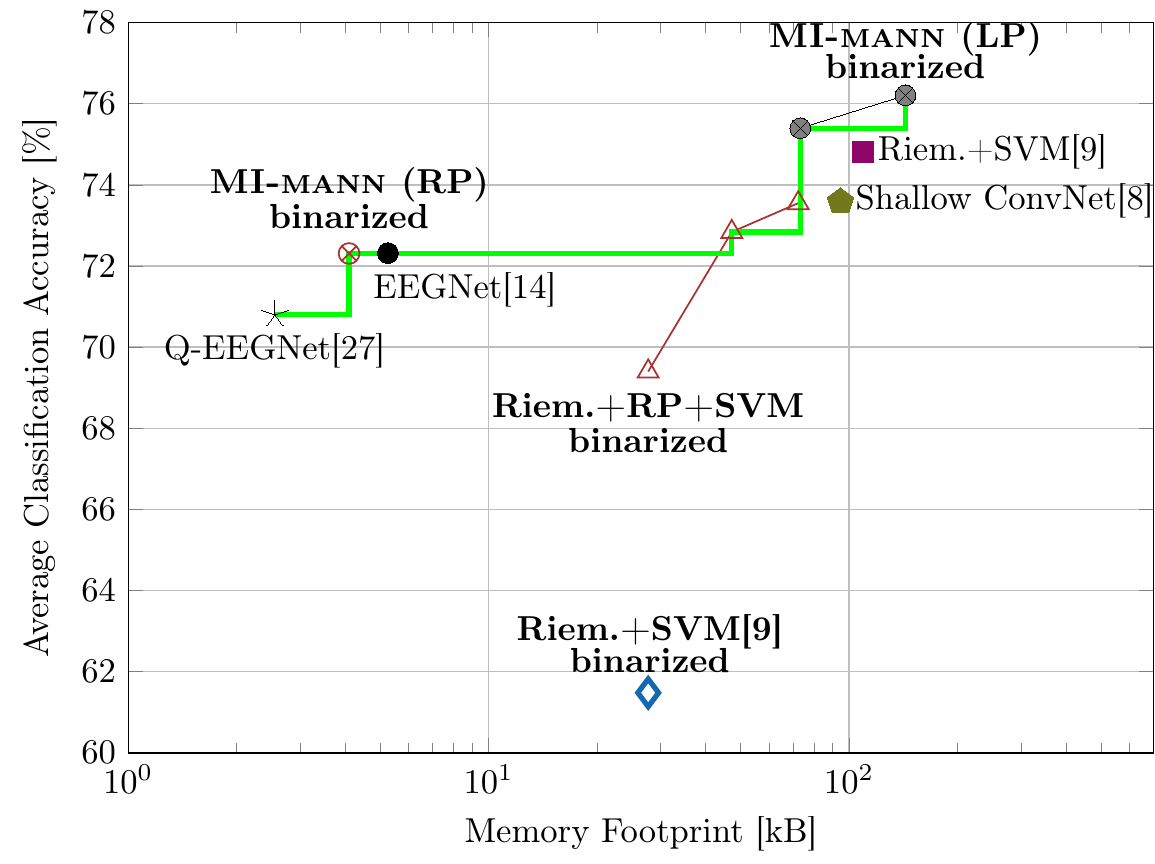}
    \caption{Main results of this work: Average classification accuracy (\%) vs. memory footprint on BCI Competition IV-2a. Our proposed binarized classifiers (bold labels) are Pareto optimal, connected with the green line. }
    \label{fig:overview_results}
\end{figure}   

Both feature-based and CNN-based approaches inherently extract a large number of real-valued features that significantly increase the number of weights and complexity of a classifier.
Such a high memory footprint and computational complexity prohibit the deployment of the model on a resource-limited device, e.g., a microcontroller, for real-time, near-sensor classification at the edge. 
One viable option is to transform those features to binary space with distance-preserving methods such as random projection~\cite{Rachkovskij2017BinaryEstimation}.
Interestingly, the weights in the matrix of random projection do not need to be stored (i.e., can be \emph{rematerialized} by a random function on the fly), or can be realized by emerging memristor~\cite{Du2017ReservoirProcessing, Mountain2018MemristorProcessor,Chakraborty2019Input-AwareDetection, Burr2016RecentTechnology,Karunaratne2020} and optical~\cite{Saade2016RandomLight} devices.
%
A readout function layer can then effectively analyze the projected features for various classification tasks, e.g., in EEG~\cite{Tan2016ApplyingBCI,Song2018SelectiveDecoding}, electrocardiography (ECG) signals~\cite{Kim2015Large-scaleHashing,Kim2016AnalysisSeries}, and electrocorticography (ECoG)~\cite{ZHANG2017BinaryMachine}.
On the other hand, for the CNN-based approaches in MI-BCIs, quantization methods to 8-bit fixed-point weights and activations are developed~\cite{Schneider2020Q-EEGNet:Interfaces}, but having a CNN model with full, or partial, binary weights is still missing in MI-BCIs.


In this paper, we extend our work in~\cite{Hersche2020BinarySVM} by proposing methods to binarize classification models for feature-based \emph{and} CNN-based MI-BCI classification approaches, summarized in Fig.~\ref{fig:overview}.
For the first approach presented in~\cite{Hersche2020BinarySVM}, we propose to embed multi-spectral, real-valued Riemannian covariance features effectively to $d$-dimensional binary Hamming space using bipolar sparse random projections. 
In the binary space, a linear SVM is trained and binarized such that classification is solely based on computationally efficient Hamming distance calculations. 
We extend~\cite{Hersche2020BinarySVM} by a second approach, where we propose to apply the concept of memory-augmented neural networks (MANNs)~\cite{Graves2014NeuralMachines,Laguna2019DesignNetworks,Karunaratne2020_MANN} for the CNN-based MI-BCI classifiers for the first time.
Inspired by~\cite{Karunaratne2020_MANN}, we replace the fully connected layer of the EEGNet with an augmented memory whose weights are binary.
Such \mimann architecture flexibly covers a wide range of classification accuracies based on the available memory: using a bipolar random projection compresses the EEGNet at the same accuracy, while using a learned projection expands the EEGNet and provides higher accuracy.

We compare the memory footprint and accuracy of our methods on the 4-class MI-dataset of the BCI competition IV-2a, summarized in Fig~\ref{fig:overview_results}.
Our binarization methods achieve Pareto-optimality with the following main results:
\begin{itemize}
    \item Randomly projecting multi-spectral Riemannian features to binary Hamming space at the same dimension as the original features, and training a binarized SVM, yields 7.92\% higher classification accuracy compared to plain binarized SVM without random projection. 
    Increasing the binary dimension to $d$=100\,000 improves the accuracy to 73.55\%, which is only 1.27\% lower than the original SVM in float16, but requires a 1.51$\times$ smaller memory footprint.
    \item Random projections enable the binarization of the augmented memory in \mimann (RP) at the same accuracy as EEGNet (72.32\%) with a 1.28$\times$ smaller memory footprint. 
    Thanks to the capability of \mimann to \emph{train} the feature extractor (i.e., EEGNet) to generate binary vectors, the dimension of the binary Hamming space could be reduced to $d$=256. 
    Additionally, allowing the projection in \mimann (LP) to be trainable, too, yields 76.21\% accuracy but increases the memory footprint by 27.28$\times$, compared to EEGNet.
    Further reducing the dimension of learned projection to $d$=128 achieves 75.40\%, which is 1.81\% more accurate and 1.29$\times$ smaller than Shallow ConvNet.
\end{itemize}

We have organized the remainder of this article as follows.
We introduce the BCI Competition IV-2a dataset and related work for MI classification of both feature-based and CNN-based approaches in Section~\ref{sec:background}. 
Section~\ref{sec:riemann} describes the proposed binarization of the classification of large multi-spectral Riemannian features using sparse bipolar random projection and binarized SVM. 
Then, in Section~\ref{sec:mann}, we present \mimann, which binarizes features in EEGNet using learned or random projections and a binary augmented memory.
In Section~\ref{sec:results}, we evaluate both feature-based and CNN-based approaches and the proposed binarized versions on the BCI Competition IV-2a according to classification accuracy, memory footprint for storing model parameters, and computational complexity in inference. 
Section~\ref{sec:conclusion} concludes the paper. 

\section{Background}\label{sec:background}
\subsection{BCI Competition IV-2a dataset}\label{sec:datasets}
The BCI Competition IV-2a dataset~\cite{Brunner2008BCIA} consists of EEG data from nine different subjects with four different MI tasks, namely the imagination of the movement of the left hand, right hand, both feet, and tongue.  
Two sessions were recorded on two different days. 
For each subject, a session consists of 72 trials per class, yielding 288 trials in total. 
One session is used for training and the other for testing exclusively. 
The signal was recorded with 22 EEG electrodes, bandpass filtered between 0.5\,Hz and 100\,Hz, and sampled with 250\,Hz. 
In addition to the 22 EEG channels, three electrooculography (EOG) channels give information about the eye movement.  
An expert marked trials containing artifacts based on the EOG signal. 
This way, 9.41\% of the trials were excluded from the dataset. The number of trials per class remains approximately balanced.

\subsection{Feature-based MI-BCI Classification}
MI is still one of the most challenging paradigms to connect the brain with an external device.
The main challenge in MI is the high variance in data between different recording sessions and different subjects; thus, most classification approaches train a separate model per subject. 
Due to the limited amount of training data per subject, traditional MI-BCIs rely on hand-crafted feature extractors and relatively simple, linear classifiers. 
EEG signals are typically pre-processed using tunable spectral and spatial filters followed by log-energy feature extraction, with filter bank common spatial pattern (FBCSP)~\cite{KaiKengAng2008FilterInterface} being the winner of the BCI Competition IV-2a and achieving an accuracy of 67\%.
The multi-spectral features are usually classified using a support vector machine (SVM), a linear discriminant analysis (LDA), or a regularized LDA~\cite{Lotte2018AUpdate}.
An alternative approach is to directly manipulate spatial EEG covariance matrices using the dedicated Riemannian geometry~\cite{Yger2017RiemannianReview,Barachant2013ClassificationApplications}.
Analogous to the FBCSP approach, the EEG signal can be divided into multiple frequency bands, where band-specific Riemannian features are calculated~\cite{Hersche2018}.  
A linear SVM on more than 32k Riemannian features, leading to overall 1.751M trainable parameters, has achieved a high classification accuracy of 75.47\%~\cite{Hersche2018} on the 4-class MI-BCI competition IV-2a dataset. 
Reducing the number of Riemannian features to 11k yields slightly lower classification accuracy of 74.82\%; however, it reduces both compute and memory requirements by 3$\times$.

\subsection{CNN-based MI-BCI Classification}
In convolutional neural networks (CNNs), the feature extractor and classifier can be combined and trained simultaneously.
While being successful in image classification, CNNs are gaining attention in MI-BCIs as well~\cite{Lotte2018AUpdate}. 
Schirrmeister et al.~\cite{Schirrmeister2017DeepVisualization} provide an elaborate study on CNN architectures for MI-BCI, where the small Shallow ConvNet achieves an accuracy of 73.59\% on the 4-class dataset. 
%
Shallow ConvNet is inspired by the classic spectral and spatial filtering with log-energy features and requires  47\,324 parameters. 

CNN++~\cite{Zhao2017OnNetworks} could further improve the accuracy to 81.1\% by proposing a much deeper network, which results in a larger model with 221\,k parameters. 
%
However, CNN++ uses not only the 22 EEG channels but also the 3 EOG channels for classification, which was not allowed in the BCI Competition IV-2a.
TPCT~\cite{Li2020AElectrodes} is the current SoA network, achieving an accuracy of 88.87\%.  %
It spatially arranges frequency band features of every EEG channel on an image according to their electrode positions and classifies them with a VGG-like CNN. 
TPCT is currently not only the most accurate CNN on the BCI Competition IV-2a, but also the largest with 7.78\,M trainable parameters. 
The large model sizes of CNN++ or TPCT prevent their deployment on portable, resource-constrained embedded devices.


On the contrary, EEGNet~\cite{Lawhern2018EEGNet:Interfaces} is a much smaller network requiring only 1716 trainable parameters.
It features a similar structure like the Shallow ConvNet.
However, it uses spatial separable convolutions and more pooling layers, which reduces the number of weights of the convolutional layer and the size of the fully connected layer. 
EEGNet enables not only the classification of MI, but also of P300 event-related potential, feedback error-related negativity, and movement-related cortical potential. 
Its flexibility and small size, however, comes at the cost of significantly lower accuracy, e.g., 67\% for 4-class MI. 
In~\cite{Uran2019ApplyingAnalysis}, EEGNet was modified by changing the pooling layers and expanding the network to 2036 trainable parameters for achieving 72\% accuracy.   
The small model size of EEGNet allows its deployment on a tightly resource-constrained embedded device, which would not be possible with larger models like CNN++ or TPCT.
In~\cite{Wang2020AnComputing}, EEGNet was applied to the large Physionet Motor Movement/Imagery Dataset~\cite{goldberger2000physiobank}, achieving SoA accuracy.
The model was ported to an ARM Cortex-M7 using \textsc{Cube.AI}, i.e., the X-CUBE-AI expansion package of STM32CubeMX. 
In Q-EEGNet~\cite{Schneider2020Q-EEGNet:Interfaces}, all weights and activations are quantized to 8-bit fixed-point using quantization-aware training, which achieved 70.8\% on 4-class MI.
 On a parallel ultra-low power (PULP) System-on-Chip~\cite{Pullini2018Mr.Processing}, the quantization as well as other hardware-aware optimizations allowed for a 252$\times$ more energy-efficient inference of EEGNet compared to the implementation on an ARM Cortex-M7.

This work exclusively studies the last classification layer's quantization in both feature-based Riemannian and CNN-based EEGNet approaches, summarized in Fig.~\ref{fig:overview}. 
In both cases, we use random (or learned) projections to map the real-valued features to the binary space.
Fixed multi-scale Riemannian features are projected to 100\,000-d Hamming space, where a linear SVM can be trained and binarized, too.
Moreover, the last layer of EEGNet is binarized with a random or learned projection, and augmented with a binary memory. 
We train the EEGNet feature extractor to generate compressed binary representations, which reduces the required binary space to $d$=256. 

\begin{figure*}[htbp]
\includegraphics[width = 1\textwidth]{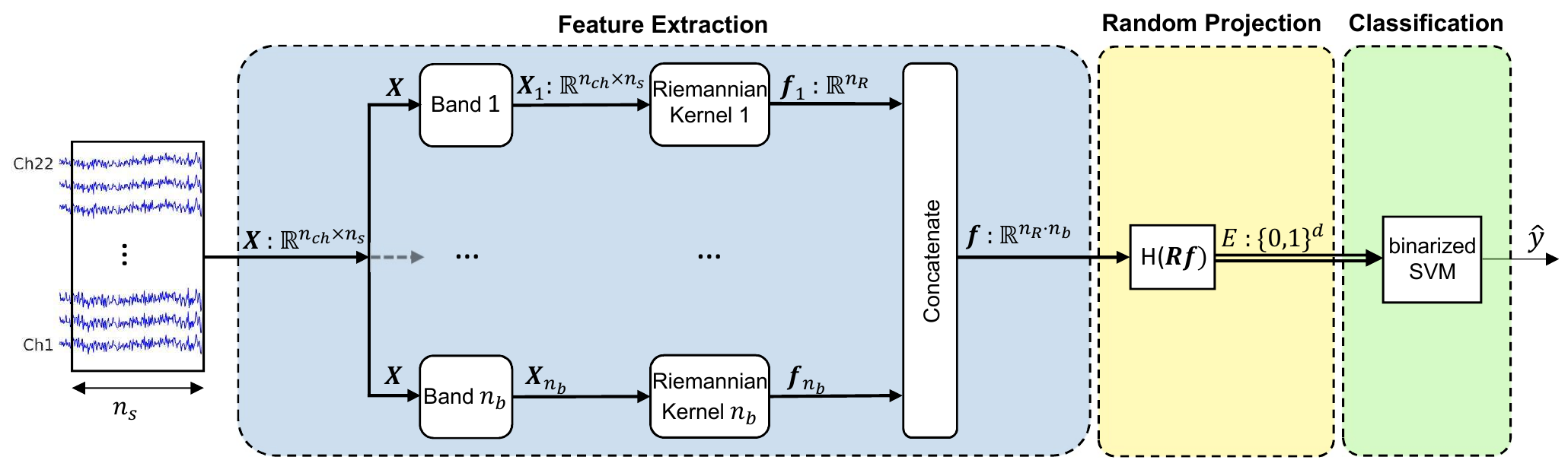}
\caption{Overall architecture for binarized learning and classification of EEG signals with feature-based classification, modified from~\cite{Hersche2020BinarySVM}. 
The EEG signal $\boldsymbol{X}$ of one temporal window with $n_s$ samples and $n_{ch}$ channels is processed at the time. 
Every EEG channel is divided into $n_b$ frequency bands ($b_1 - b_{n_b}$) using second order Butterworth band pass filters. 
A Riemannian covariance kernel computes spatial energy features which are concatenated and binarized using sparse random projection. 
Binary features $E$ are classified with a binarized SVM.}
\label{fig:architecture}
\end{figure*}
\section{Binarizing Riemannian Features with Sparse Bipolar Random Projections}\label{sec:riemann}

This section presents the first main contribution of the paper, which is to binarize multi-spectral Riemannian features with sparse bipolar random projections, and classify them with a binarized SVM, shown in Fig.~\ref{fig:architecture}. 

\subsection{Riemannian Covariance Features}\label{subsec:feat_extr}
%
%
We use a recent approach~\cite{Yger2017RiemannianReview}, which extracts features from EEG by directly manipulating spatial EEG covariance matrices using the dedicated Riemannian geometry.
First, we estimate the covariance matrix $\mathbf{C}:=\mathbf{C}^{(i)}$ of a trial $i$ from the multi-channel EEG signal $\mathbf{X}:=\mathbf{X}^{(i)}\in \Reals^{n_{ch}\times{n_s}}$ with $n_{ch}$ channels and $n_s$ samples:
\begin{align}
\boldsymbol{C} = \frac{1}{n_s-1} (\boldsymbol{X} \boldsymbol{X}^T + \alpha \boldsymbol{I}_{n_{ch}}), 
\end{align}
where $\boldsymbol{I}_{n_{ch}}$ is the $n_{ch} \times n_{ch}$ identity matrix and $\alpha$ a regularization constant ensuring positive definiteness of the estimated covariance matrices set to $0.1$.
The Riemannian kernel $\boldsymbol{f}$ calculates $n_R = n_{ch}(n_{ch}+1)/2$ output features based on the input covariance matrix $\mathbf{C}$: 
\begin{align}
K: \Reals^{n_{ch} \times n_{ch}} \rightarrow \Reals^{n_R}, 
\end{align}
and is defined as
\begin{align}
K\left(\boldsymbol{C}\right) = \textrm{vect}\left(\textrm{logm}\left(\boldsymbol{C}_{ref}^{-1/2}\boldsymbol{C}\boldsymbol{C}_{ref}^{-1/2} \right) \right), \label{eq:riemann_feat}
\end{align}
where logm(.) is the matrix logarithm and vect(.) the $\ell_2$-norm preserving half vectorization of a matrix~\cite{Barachant2013ClassificationApplications}. 
The computation of the matrix logarithm involves the eigenvalue decomposition (EVD), the logarithm computation on the eigenvalues, and the back transformation. 
The EVD can be efficiently divided into a Housholder transformation~\cite{burden1997numerical} for tridiagonalization and an iterative QR-decomposition using an implicit Wilkinson shift~\cite{wilkinson2013linear}.
The reference covariance matrix $\mathbf{C}_{ref}$ is the geometric mean over all covariance matrices in the training set~\cite{Moakher2005AMatrices}.
The Riemannian kernel does not need labeled data and is therefore unsupervised. 
The multiplication of the covariance matrix $\mathbf{C}$ with $\mathbf{C}_{ref}^{-1/2}$ on both sides is interpreted as spatial whitening of $\mathbf{C}$.

In analogy to frequency band common spatial pattern (FBCSP), the set of Riemannian features is extended to multi-spectral features by using multiple Riemannian kernels on different frequency bands of the multi-channel EEG signal. 
The signal is divided into multiple frequency bands using a filter bank. 
A separate Riemannian kernel is used with $\mathbf{C}_{ref}$ computed solely on the corresponding frequency band. 
A recent work~\cite{Hersche2018} with high classification accuracy suggests using $n_b$=43 overlapping frequency bands within the 4--40\,Hz band with bandwidths varying between 2--32\,Hz. 

We apply the multi-spectral Riemannian feature extractor on the BCI Competition IV-2a by extracting EEG recording from 3.5\,s ($n_s$=875), starting at 0.5\,s after the MI cue according to the timing scheme of the competition.
We use all 22 EEG channels, which yields $n_R$=253 features per frequency band and a total of $n_R\cdot n_b$=10\,879 multi-spectral Riemannian features. 

\subsection{Sparse Bipolar Random Projection}
An embedding is a representation for which the computation of distances directly gives an estimate of the distances in their initial representation~\cite{Rachkovskij2017BinaryEstimation}. 
The building of such representations is provided by binary locality-sensitive hashing (LSH) functions,  which ensure that similar elements are statistically likely to be embedded into the same value~\cite{Kim2015Large-scaleHashing}. 
Once mapped to the binary Hamming space, the similarity is computed with the Hamming distance. 

Here, we use random projections to embed real-valued feature vectors to the binary d-dimensional Hamming space $\Binary^d:=\lbrace0,1\rbrace ^ d$. 
Random projections are usually used for dimensionality reduction in the Euclidean space~\cite{Bingham2001RandomReduction}. 
The Johnson-Lindenstrauss lemma~\cite{Johnson1984ExtensionsSpace} ensures distances between two points in the projected space to be preserved if the output dimension is suitably high. 
Such projections deal with embeddings between Euclidean spaces.
However, here the data is projected to a high-dimensional Hamming space.
Recently, it has been shown~\cite{Rachkovskij2017BinaryEstimation} that random projections can indeed project data to a high-dimensional Hamming space while preserving the distance between points with success in monitoring arterial blood pressure via ECG signals~\cite{Kim2015Large-scaleHashing,Kim2016AnalysisSeries}.

Random projection to binary space is defined as 
\begin{align}
   E = \textrm{H}(\boldsymbol{R} \boldsymbol{f}), 
\end{align}
where H(.) is the component-wise Heavyside step function 
\begin{align}
\label{eq:heavyside}
 \textrm{H}(\mathbf{z}[i])  = 
  \begin{cases}
    1  & \textrm{if } \mathbf{z}[i] \geq 0 \\
    0  & \textrm{if } \mathbf{z}[i] < 0,   \\
  \end{cases}
\end{align}
and $\boldsymbol{R} \in \Reals ^{d \times n_f}$ the projection matrix \cite{Rachkovskij2017BinaryEstimation}.
Usually, the components $r_{i,j}$ in $\boldsymbol{R}$ are drawn from an i.i.d. Gaussian normal distribution ($r_{i,j} \sim \mathcal{N}(0,1)$). 
However, the Gaussian projection matrix can be replaced by a much simpler one such as the sparse bipolar random matrix \cite{Achlioptas2001Database-friendlyProjections}:
\begin{align}
 r_{i,j} = 
   \begin{cases}
    +1  & \textrm{with probability } \frac{1-s}{2} \\
    0  & \textrm{with probability } s \\
    -1  & \textrm{with probability } \frac{1-s}{2}, \\
   \end{cases}
\end{align}
where $s \in [0,1]$ is the sparsity, i.e., the number of zero elements divided by the total number of elements.
Achlioptas \cite{Achlioptas2001Database-friendlyProjections} has shown that by using a sparsity of $s=1/3$ this projection comes without any sacrifice in the quality of embedding compared to the plain Gaussian projection. 
In this application, the use of a bipolar instead of a Gaussian projection matrix yielded no loss in accuracy; furthermore, we could use projection matrices with a sparsity of $s=9/10$ without losing performance. 
The use of projection matrices, which are both bipolar and sparse, reduces the computational complexity of projection: the bipolarity limits the dot product to a sequence of additions and subtractions, while the sparsity reduces the number of operations.
Random entries of the projection matrix do not need to be stored permanently, but can be efficiently \emph{regenerated during operation} with a random number generator. 
This process is also known as \emph{rematerialization} that is repeatable and requires an arbitrary seed, which requires negligible 32-bit storage.
Thus, the use of random projections is not increasing the memory footprint for storing a model on an embedded device~\cite{Hersche2018ExploringInterfaces}.

\subsection{Binarized SVM}\label{subsec:bin_svm}
This section describes how a linear SVM is binarized to do binary inference solely based on Hamming distance computations in projected d-dimensional space. 
We use the fact that there exists a one-to-one mapping between the cosine similarity and the normalized Hamming distance of two bipolar vectors: 
\begin{align}
    \frac{<\mathbf{a},\mathbf{b}>}{||a||_2 ||b||_2} &= \frac{1}{d} \sum_{i=1
}^d \mathbf{a}[i] \cdot \mathbf{b}[i] \\
&=\frac{1}{d}\left(d+\sum_{i=1}^d\left(\mathbf{a}[i] \cdot \mathbf{b}[i] -1\right) \right) \\
&=\frac{1}{d}\left(d+\sum_{i=1}^d 2\left(-\mathbf{1}_{\mathbf{a}[i] \neq \mathbf{b}[i]}\right)\right) \\
&=1 - 2\mathrm{d_h}(\mathbf{a},\mathbf{b}),
\end{align}
where $<.,.>$ is the inner product, $||.||_2$ the $\ell_2$ norm, and $\textrm{d}_h(.)$ the normalized Hamming distance. 
As a consequence, we use binary and bipolar representations interchangeably, e.g., training a model on bipolar vectors and execute inference on binary vectors using the Hamming distance. 

When neglecting some scaling factors, the decision function of the original linear SVM without bias relies on cosine similarity and is defined as  
\begin{align}
    \hat{y}= \operatornamewithlimits{\textrm{argmax}}_{i=1,...,n_{cl}}<\mathbf{w}_i,\mathbf{f}>,
\end{align}
where $\mathbf{w}_i \in \Reals^d$ is the learned support vectors of class $i$ with unit norm.
For training the linear SVM---still in full float32 precision---on binary features, we map all elements in $E\in\Binary^d$ to bipolar values $\lbrace-1,1\rbrace$. 
The learned support vectors are then binarized using the component-wise Heavyside step function:
\begin{align}
    W_i=\textrm{H}(\textbf{w}_i) \quad i=1,...,n_{cl}
\end{align}
During inference, the binarized SVM classifies a binary vector $E$ by searching for the binary support vector with smallest Hamming distance to $E$: 
\begin{align}
    \hat{y}= \operatornamewithlimits{\textrm{argmin}}_{i=1,...,n_{cl}}\textrm{d}_h(W_i,E).
\end{align}

\section{\mimann: Learning Compact Binary Representations with Memory-augmented Neural Networks}\label{sec:mann}
%
One main challenge traditional neural networks face is the inability to recognize new classes without complete retraining~\cite{Laguna2019DesignNetworks}. 
Retraining on samples of a new, unseen class often yields to large performance degradation in recognizing the ``old'' classes, also known as catastrophic forgetting~\cite{McCloskey1989CatastrophicProblem}.  
To address this challenge, memory-augmented neural networks (MANNs) add an external memory, which can easily be updated or extended without retraining the entire model~\cite{Graves2014NeuralMachines,Laguna2019DesignNetworks}.  
MANNs have been proven to be particularly useful in few-shot learning problems, such as the Omniglot task containing a large number of 1623 characters with only 20 samples per character~\cite{Lake2015Human-levelInduction.}. 
%

Such high numbers of classes are not encountered in MI-BCIs; however, augmenting an MI-BCI model with an external memory allows to update/extend the model, e.g., 
\begin{itemize}
    \item Adding a new MI class without retraining the whole model.
    \item Calibrating the model at the beginning of a new \emph{session} to mitigate high inter-session variance in EEG.
    \item Calibrating the model on a new, unseen \emph{subject} due to high inter-subject variance. 
\end{itemize}
In a nutshell, this novel architecture could quickly store new information in the external memory, and adapt to a changing environment typical of BCIs.

Classification in MANNs requires mostly $\ell_2$-distance computation between a query vector and all entries in the external memory, where the complexity grows with the size of the memory.
Efforts have been invested in simplifying the computation by using alternative distance metrics such as $\ell_1$ or $\ell_\infty$~\cite{Laguna2019DesignNetworks}. 
More recently, a MANN has been proposed which is trained to generate high-dimensional bipolar (or binary) vectors by construction~\cite{Karunaratne2020_MANN}. 

Here, we present \mimann, a binarized MANN for MI-classification, which augments EEGNet with a projection layer for binarization of the features and an external binary memory. 
To best of our knowledge, this is the first application of MANNs in the MI-BCI context. 
%

\begin{figure}
\centering
\footnotesize
\includegraphics[width=\linewidth]{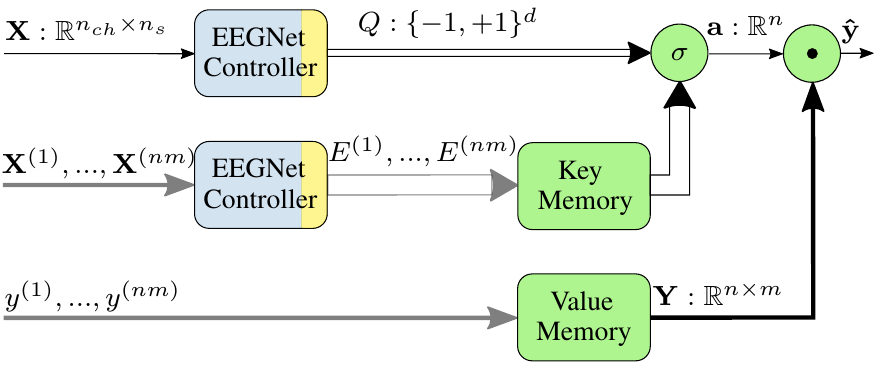}
\caption{\mimann architecture. Key and value memory are filled with processed samples $\mathbf{X}^{(1)},...,\mathbf{X}^{(nm)}$ and values $y^{(1)},...,y^{(nm)}$ from the support-set.
Query $\mathbf{X}$ is processed by the EEGNet controller, the attention function $\sigma$, and classified with matrix-vector product.
}
\label{fig:memaugm}
\end{figure}

\subsection{\mimann}
\begin{figure*}
\centering
\footnotesize
\includegraphics[width=\textwidth]{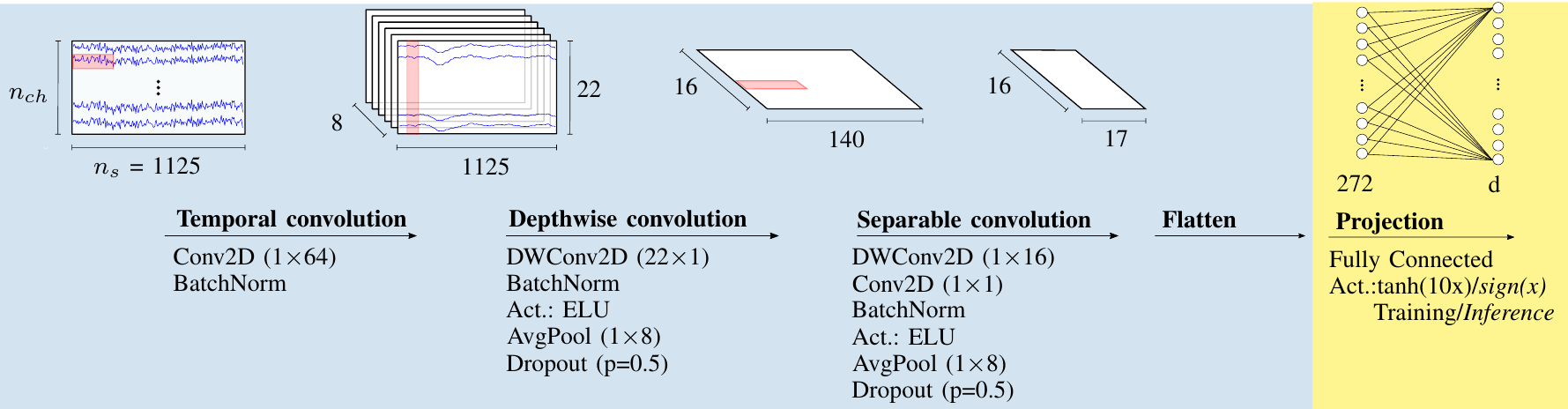}
\caption{EEGNet controller. Convolutional layers of EEGNet~\cite{Lawhern2018EEGNet:Interfaces} are extended with a projection layer.}
\label{fig:eegnet_controller}
\end{figure*}

Fig.~\ref{fig:memaugm} depicts the proposed architecture of \mimann, consisting of an EEGNet controller, a key memory, a value memory, as well as an attention function $\sigma$. 
Here, we describe the entire functionality using bipolar vectors and cosine similarities as it is used in training. 
In inference, however, all binary blocks (key memory, value memory, attention) are implemented with binary vectors using Hamming distance according to the distance mapping described in Section~\ref{subsec:bin_svm}.
The EEGNet controller is responsible for extracting $d$ distinctive bipolar features from the input signal $\mathbf{X}$. 
In a first step, we assume that the EEGNet controller has already been trained; the training procedure will be explained in Section~\ref{subsec:mat_controller}. 
%

Our MANN system has two memories: the key memory and the value memory.
At training time, a subset of the training set is chosen to be the so-called \emph{support-set}.
Each example $\left(\mathbf{X}^{(i)}, y^{(i)}\right)$ in the support-set is then pre-processed and stored in the two components of the memory:
the input $\mathbf{X}^{(i)}$ is passed through the EEGNet controller, obtaining a vector with bipolar components $E$ which is stored into the \emph{key memory};
simultaneously, the label $y^{(i)}$ is one-hot encoded (remember that MI is a classification task) and stored into the \emph{value memory}.
In MANNs, the number of classes is referred to as \emph{ways} and the training examples to \emph{shots}. 
An m-way/n-shot classifier is provided with $m\cdot n$ samples $\mathbf{X}^{(1)},\mathbf{X}^{(2)},...\mathbf{X}^{(mn)}$ and values $y^{(1)},...,y^{(mn)}$ from the support-set.

A query $\mathbf{X}$ is encoded into $Q$ using the EEGNet controller and passed through the attention function $\sigma$.
The attention function computes the cosine similarity between the encoded vector $Q$ and all entries in the key memory:
\begin{align}
     \mathbf{\alpha}_i:=\alpha (Q,E_i) = \frac{<Q,E_i>}{||Q||_2 ||E_i||_2} \quad i \in \lbrace1,2,...,mn\rbrace,
 \end{align}
followed by a \emph{soft absolute (softabs)} sharpening function~\cite{Karunaratne2020_MANN} yielding the attention vector $\mathbf{a}$.
Softabs is similar to softmax but relaxes the optimization constraint by obeying the attention conditions provided by high-dimensional computing; it forces the controller to generate orthogonal vectors instead of anticorrelating vectors for different classes~\cite{Karunaratne2020_MANN}.
%
The attention vector $\mathbf{a}$ is finally multiplied with the one-hot encoded training labels. 
The estimated MI class is the argmax of $\mathbf{\hat{y}}$.

We terminate this section with a simple classification example of a 2-way/2-shot classifier.
The label support-set contains the examples of $y^{(1)}=0$, $y^{(2)}=1$ ,$y^{(3)}=1$, and $y^{(4)}=0$, which are one-hot encoded and written into the value memory:
\begin{align}
    \mathbf{Y} = \left( \begin{matrix} 1 & 0 \\ 0 & 1 \\ 0 & 1 \\ 1 & 0 \end{matrix} \right).
\end{align}
Similarly, the encoded feature vectors $E^{(1)},E^{(2)},E^{(3)}$, and $E^{(4)}$ are written into the key memory. 
For classifying an attention vector, e.g., $\mathbf{a}=(0.2,0.3,0.4,0.1)$, we compute $\mathbf{\hat{y}}=\mathbf{a}\mathbf{Y}=(0.3,0.7)$. 
The estimated MI class would be $\hat{y}=1$.

\subsection{EEGNet Controller for Generating Bipolar Features with Random or Learned Projections}

We propose to generate bipolar features using an EEGNet controller (see Fig~\ref{fig:eegnet_controller}), which resembles the convolutional layers of EEGNet, without the fully connected classification layer, and a projection layer.
We first extract EEG recording from 4.5\,s ($n_s$=1125), starting 0.5\,s before the cue. 
The temporal convolution block filters the EEG data in the time domain, before the channels are combined with spatial filters in the depthwise convolution. 
The ELU activation after the separable convolution makes the generation of bipolar feature values hard. 
Therefore, we introduce a projection layer to dimension $d$ with a sharpened tanh(10x) activation.
The steep activation function ensures almost bipolar output features. 
We use the tanh activation in training for backpropagating the gradients, as it keeps the controller differentiable. 
In inference, the activation is replaced by the sign function in the bipolar case, or the Heaviside step function in the binary case. 

We distinguish between random and learned projections. 
The random projection is initialized with bipolar, \emph{dense} values at the beginning of training and fixed from thereon. 
Best training results were achieved when scaling all entries according to the maximum value of Xavier's uniform initialization~\cite{Glorot2010UnderstandingNetworks}. 
This scaling factor can be efficiently embedded into the batch norm layer to save compute efforts. 
The values of the random projection matrix can be generated on the fly with a random number generator; therefore, it does not require additional memory for storage. 
In contrast, the learned projection is implemented as a trainable, fully connected layer without a bias. 
The learned projection significantly adds model parameters, which need to be stored on the device. 
However, it also adds the capability to learn more distinctive representations.

\subsection{Training the EEGNet Controller}\label{subsec:mat_controller}
This section describes the training procedure of the EEGNet controller, which is dominated by teaching the controller how to learn from a few examples.
The training alternately updates either the key-value memory \emph{or} the controller weights.
We exclusively use samples $\left( \mathbf{X}^{(i)},y^{(i)}\right)$ from the training set. 

In the beginning, the controller weights are initialized randomly with uniform distribution. 
In an initialization step, we randomly choose a support-set of $mn$ samples $\mathbf{X}^{(1)},\mathbf{X}^{(2)},...,\mathbf{X}^{(mn)}$, pass them through the (random) controller, and store the feature vectors $E^{(1)},E^{(2)},...,E^{(mn)}$ into the key memory.
At this stage, the key memory is not guaranteed to be bipolar yet; the bipolarization will follow in the last stage of the training.
The corresponding labels $y^{(1)},y^{(2)},...,y^{(mn)}$ are one-hot encoded and stored into the value memory.

After the initialization phase, the controller and key-value memory are updated iteratively, where one training epoch is defined as follows: 
\begin{enumerate}
   \item \textbf{Update the controller.} 
   We first pick a random set of batch size $k$ samples and pass them through the network, resulting in the estimated probability distribution of all samples $\mathbf{\hat{y}}^{(1)},\mathbf{\hat{y}}^{(2)},...,\mathbf{\hat{y}}^{(k)}$. 
    Next, the binary cross-entropy (BCE) loss is computed and backpropagated through the network. 
    The controller weights are finally updated using Adam's optimizer. 
    In this stage, the key-value memory remains fixed.
    \item \textbf{Update the key-value memory.}
    After adjusting the controller, the key-value memory is entirely re-written with new samples, which were not used in the training of the controller.
\end{enumerate}
After every epoch, the training data is shuffled such that the samples for training the controller and updating the key-value memory change.
During inference, the activation function of the projection layer (tanh) is replaced by the sign or Heaviside function to generate bipolar or binary vectors. 
Similarly, the key-memory is bipolarized or binarized. 

\begin{figure}[htbp]
\centering
\footnotesize
\includegraphics[width=\linewidth]{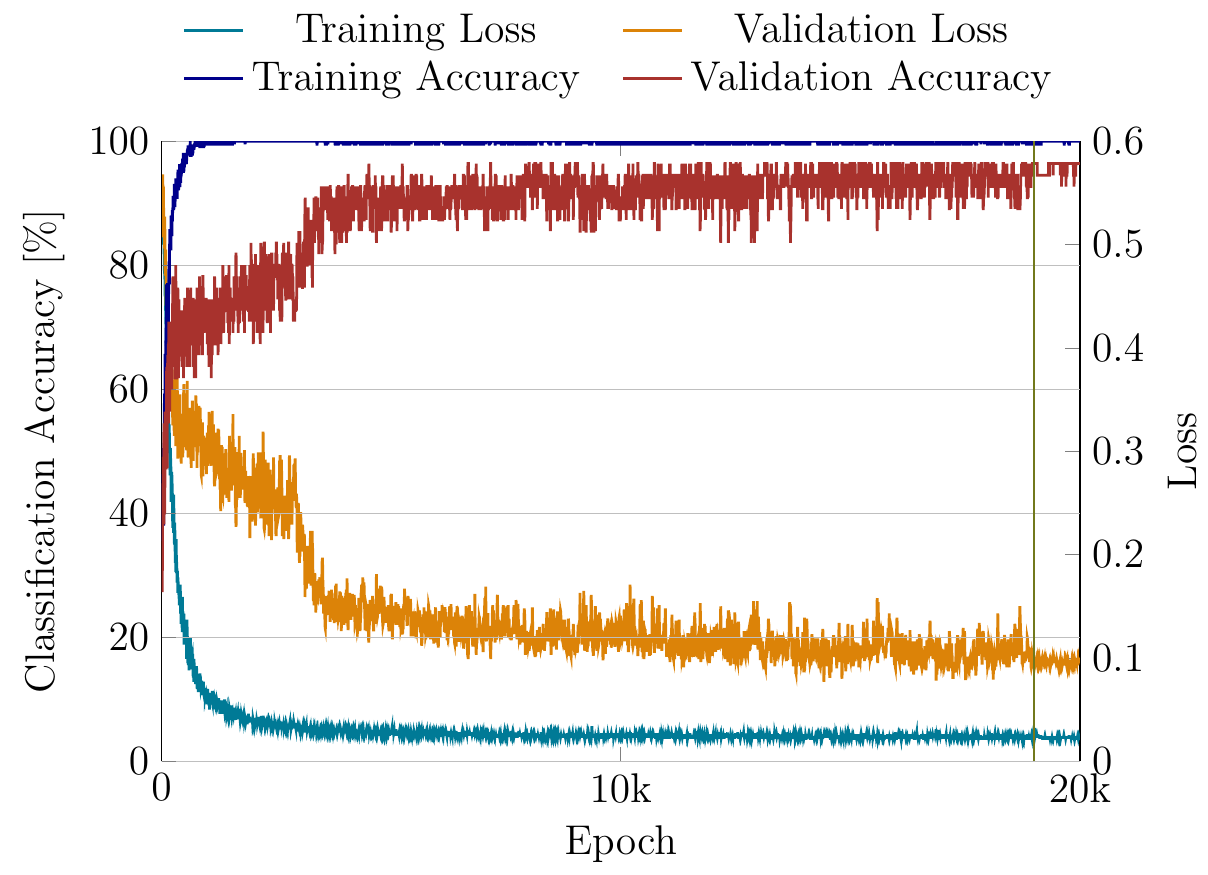}
\caption{Accuracy and loss on training data of subject 7 of BCI Competition IV-2a dataset. 
The binary \mimann with learned projection ($d$=256) is trained with learning rate 1e-3 for 19\,000 epochs, and with 1e-4 for the last 1000 epochs. 
}
\label{fig:learning_memaugm}
\end{figure}

Fig.~\ref{fig:learning_memaugm} shows the accuracy and loss on training (80\%) and validation (20\%) data of the training set of subject 7 of the BCI Competition IV-2a dataset. 
The network is trained for 20\,000 epochs using a batch size of 64 and learning rate 1e-3. 
Even though the model achieves a training accuracy of almost 100\% after $\approx$1000 epochs, it still improves on the validation data when continuing with training. 
We see a high variance in classification accuracy and loss on the validation data; therefore, the learning rate is reduced to 1e-4 for the last 1000 epochs.
%

\begin{table*}[htpb]
\small
\centering
\caption{Classification accuracy (\%) on 4-class MI dataset of BCI Competition IV-2a using Riemannian and CNN-based approaches. 
}
\label{tab:results}
\begin{threeparttable}
\begin{tabular}{lccccccccc} 
\toprule
& \multicolumn{5}{c}{Riemannian} &\multicolumn{4}{c}{CNN} \\ 
\cmidrule(r){1-1} \cmidrule(r){2-6} \cmidrule(r){7-10} 
 & SVM & LDA & SVM & LDA & SVM & EEGNet & \mimann & \mimann & \mimann \\ 
 & float16 & float16 & binarized & binarized & binarized & float16 & binarized & binarized & binarized \\ 
 \cmidrule(r){2-6} \cmidrule(r){7-10} 
 Projection & - & - & - & - & sparse RP & - & - & RP & LP \\ 
 d 	&	10\,879 &	10\,879 &	10\,879 &	 10\,879 &	 100\,000 &	 272& 	 272 &	256&	256 \\ 
 \cmidrule(r){1-1} \cmidrule(r){2-6} \cmidrule(r){7-10} 
 S1 &	91.81 &	88.26 &	78.65 &	78.29 &	90.46 &	84.36 &	75.40 &	81.47 &	82.24\\ 
S2 &	51.59 &	58.66 &	45.58 &	44.88 &	53.96 &	54.06 &	45.32 &	57.68 &	64.66\\ 
S3 &	83.52 &	82.78 &	68.13 &	71.79 &	79.16 &	87.91 &	85.81 &	90.82 &	93.19\\ 
S4 &	73.25 &	53.51 &	57.89 &	56.58 &	71.49 &	63.16 &	54.06 &	60.28 &	60.83\\ 
S5 &	63.41 &	59.42 &	42.03 &	40.94 &	65.18 &	67.39 &	52.26 &	62.92 &	74.57\\ 
S6 &	59.07 &	57.21 &	47.91 &	50.23 &	56.98 &	54.88 &	49.24 &	52.01 &	57.34\\ 
S7 &	86.64 &	89.53 &	71.12 &	73.29 &	82.42 &	88.09 &	79.11 &	86.34 &	88.56\\ 
S8 &	81.55 &	81.92 &	71.59 &	75.65 &	79.63 &	76.75 &	79.11 &	82.12 &	83.41\\ 
S9 &	82.58 &	77.65 &	70.45 &	73.86 &	82.65 &	74.24 &	71.83 &	77.14 &	81.08\\ 
\cmidrule(r){1-1} \cmidrule(r){2-6} \cmidrule(r){7-10} 
 \textbf{{Avg.}} &	 \textbf{74.82} &	 \textbf{72.10} &	 \textbf{61.48} &	 \textbf{62.83} &	 \textbf{73.55} &	 \textbf{72.32} &	 \textbf{65.79} &	 \textbf{72.31} &	 \textbf{76.21}\\ 
Std. &	12.37 &	13.10 &	12.01 &	13.09 &	11.17 &	11.86 &	13.78 &	12.66 &	11.32\\ 
p-value$^*$ &	 - &	 0.260 &	 0.008 &	 0.008 &	 0.214 &	 - &	 0.015 &	 0.594 &	 0.038\\ 
\bottomrule
\end{tabular}
\begin{tablenotes}\footnotesize
\item $^*$Significance of a Wilcoxon signed-rank test with respect to baseline classifier, which is linear SVM in float16 precision for Riemannian and EEGNet for CNN-based.
\end{tablenotes}
\end{threeparttable}
\end{table*}
\section{Experimental Results}\label{sec:results}
In this section, we assess the proposed methods on the 4-class MI dataset from the BCI competition IV-2a.
In all experiments, we train a separate model (feature extractor and classifier) per subject on the training set; the test set is neither touched for training nor for validating model hyperparameters. 
%
CNN-based models are trained on an Nvidia GTX 1080 Ti GPU using PyTorch (version 1.4.0).

We measure the classification accuracy as the ratio between correct classified trials over the total number of trials. 
Moreover, we compare the memory footprint of the models for storing the learned parameters. 
Finally, we assess the computational complexity in inference of the CNN-based classifier by counting the number of multiply-accumulate (MAC) operations.

\subsection{MI Classification on Binarized Riemannian Features}
We first consider the Riemannian features and its binarization. 
An $\ell_2$-regularized linear SVM performed best on the 4-class dataset with multi-spectral Riemannian features~\cite{Hersche2018} and serves as a baseline classifier.
An LDA with automatic shrinkage, commonly used in EEG classification~\cite{Lotte2011RegularizingAlgorithms}, is used as a second baseline.
The Riemannian columns of Table~\ref{tab:results} compare the classification accuracy for float16 precision linear SVM and LDA with different binary classifiers.
The linear SVM achieves 74.82\% average classification accuracy; slightly lower results are observed with LDA at 72.10\% accuracy.

In a first step, the baseline classifier and features are binarized in their original space applying the Heavyside step function directly on features and support vectors. 
This results in a significant loss of 13.34\% in accuracy for binarized SVM and 9.27\% for binarized LDA, relative to their corresponding float16 classifier. 
However, this performance loss due to binarized classification can be recovered when applying our proposed method using sparse bipolar random projection to binary Hamming space and binarized SVM. 
We observed just a minor accuracy degradation between our RP+SVM approach and an SVM at FP16 precision (73.55\% vs. 74.82\%), which is largely compensated for by the memory saving. 


\subsection{MI Classification in Binarized \mimann}
Next, we discuss the classification accuracy of the binarized 8-shot/4-way \mimann, shown in Table~\ref{tab:results} in the CNN columns. 
Original EEGNet in float16 precision serves as a baseline, achieving an accuracy of 72.32\%. 
In a first experiment, we assess the accuracy of \mimann without using a projection layer in the EEGNet controller. 
For doing so, the activation in the separable convolution block is changed from ELU to tanh. 
Akin to the previous experiment, where Riemannian features were binarized without using random projections, we observe a significantly lower classification accuracy of 65.79\%.
This drop in accuracy is mitigated by the introduction of bipolar random projections, where we achieve almost the same accuracy of 72.31\% as full precision EEGNet at binary dimension $d$=256. 
When relaxing the constraints in the EEGNet controller and allowing learned projections, the accuracy can even be increased to 76.21\%, which is 3.89\% and 1.39\% more accurate than full precision EEGNet and  Riemannian with SVM, respectively.  

\subsection{Memory footprint}
\begin{figure*}[htbp]
\centering
\includegraphics[width=0.9\textwidth]{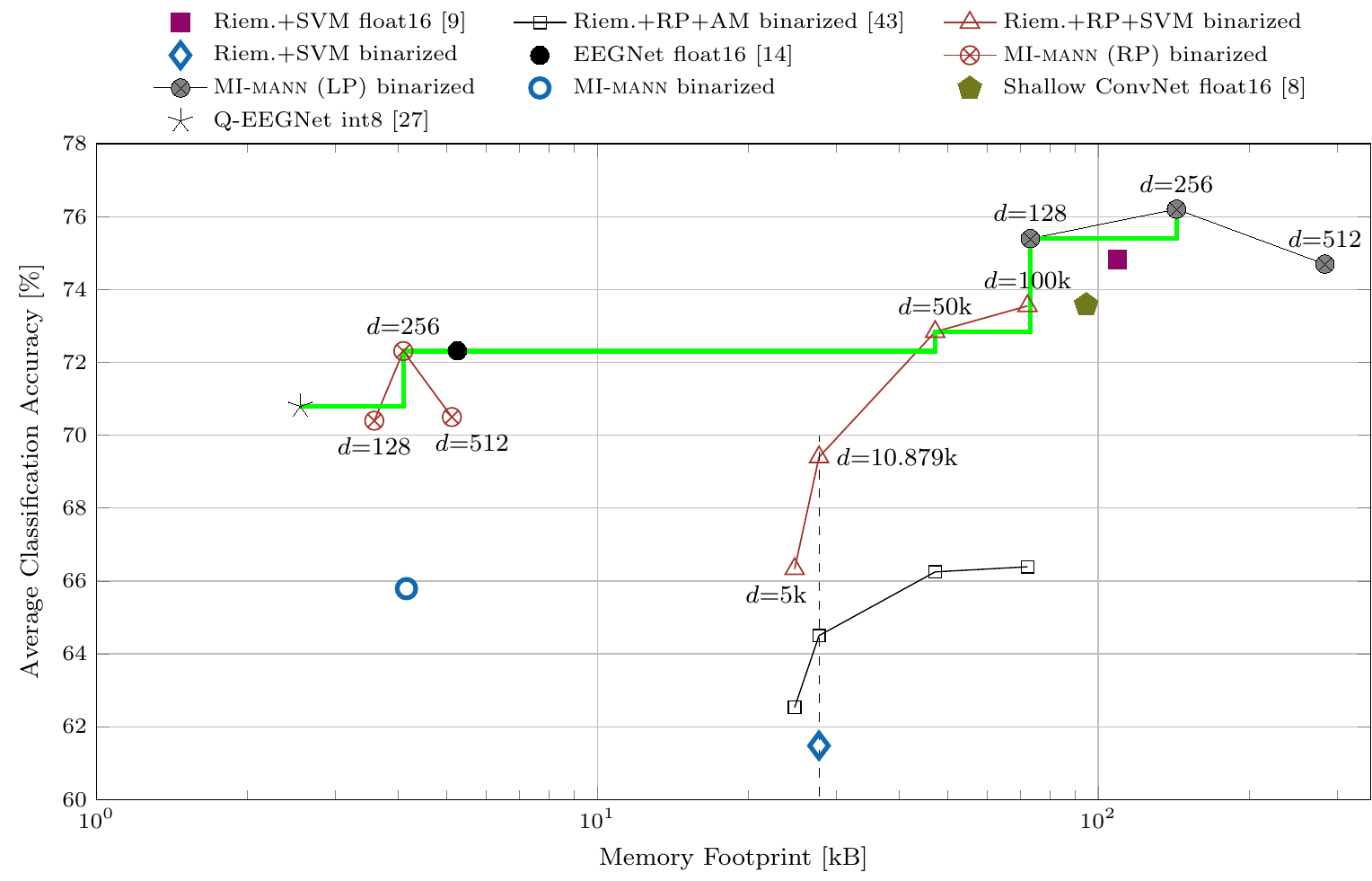}%
\caption{Average classification accuracy (\%) vs. memory footprint on 4-class MI of BCI Competition IV-2a.
Pareto-optimal classifiers are connected with a green, solid line.  
%
}
\label{fig:hd_map_results}
\end{figure*}
Fig.~\ref{fig:hd_map_results} compares the performance of all classifiers, considering not only the classification accuracy but also the memory footprint required to store learned parameters of the whole model.
Here, we include another binarized classifier~\cite{Hersche2018ExploringInterfaces} which uses random projections on Riemannian features as well, but encodes the projected binary vectors per frequency band using holographic superposition (RP+AM binarized).
The binary vectors are classified using an associative memory (AM).   
Moreover, we consider reasonable sized CNNs, which are deployable on a typical low-power microcontroller featuring a few MB of Flash memory, such as Shallow ConvNet~\cite{Schirrmeister2017DeepVisualization} with 47\,324 float16 parameters and Q-EEGNet~\cite{Schneider2020Q-EEGNet:Interfaces} with 2036 int8 parameters.
However, Fig~\ref{fig:hd_map_results} does neither include TPCT with 88.87\% accuracy due to its high memory footprint of 15.56\,MB, nor CNN++ because it violates the rules of the BCI Competition IV-2a. 
A more detailed listing of all CNN-based classifiers is available in Table~\ref{tab:comp}.

%
First, we consider the binarization of the Riemannian features.
The output dimension of the random projection is varied between $d$=5k--100k, which has a direct impact on the required memory footprint. 
Generally, the accuracy of the binarized classifier improves significantly in higher dimensions, especially when using the proposed binarized SVM readout. 
When fixing the dimension to the number of Riemannian features (i.e., $d$=10\,879 or memory footprint of $27.71$\,kB), the simple SVM binarized achieves lower accuracy compared to both RP methods. 
This supports the necessity of RP when doing binarized classification. 
At memory footprint of 72.27\,kB ($d$=100k), Riemannian+RP+SVM binarized achieves 73.55\% accuracy, which is 1.27\% lower than float16 SVM, but it requires 1.51$\times$ lower memory footprint. 
Compared to Shallow ConvNet with 73.59\% accuracy and 94.65\,kB memory footprint, RP+SVM binarized reduces the memory footprint by 1.31$\times$ at the same accuracy. 

Next, the dimension of the random projections in the binarized \mimann is varied from $d$=128--512. 
We find the optimal dimension to be $d$=256, which is closest to the number of features in EEGNet (272). 
The binarized \mimann requires 4.10\,kB memory footprint at d=$256$, which is 1.28$\times$ lower than EEGNet in float16 precision at the same accuracy.
Q-EEGNet requires the lowest memory footprint of 2.55\,kB, but also achieves with 70.8\% a lower accuracy than both EEGNet in float16 and binarized \mimann at $d$=256. 

Similar trends are observed when allowing the projection to the binary space to be trained (\mimann (LP) binarized).
Also here, the highest accuracy of 76.21\% is achieved at $d$=256. 
The use of learned projections adds a significant amount of memory: it increases the memory footprint by 13.9--54.0$\times$ for $d$=128--512, compared to the original EEGNet in float16 precision. 
However, at the lowest dimension $d$=128, the binarized memory-augmented network achieves an accuracy of 75.4\% at 73.22\,kB memory footprint, which is a reduction of the memory footprint by 1.29$\times$ and 1.48$\times$ compared to Shallow ConvNet and Riemannian+SVM float16, respectively.

To sum up, our proposed binarization methods are able to reduce the memory footprint on both the feature-based and CNN-based classifiers while maintaining similar accuracy. 
As a result, all binarized classifiers achieve Pareto optimality, shown by the green, solid line in Fig.~\ref{fig:hd_map_results}.


\subsection{Complexity of Inference}
This section discusses the complexity of classifiers during inference by counting the number of MAC operations for computing one classification, shown in Table~\ref{tab:comp}.
The Hamming distance computation for classification of binary query vectors in binarized \mimann can be implemented with bit-level operations (XOR+POPCOUNT); thus, we count the computation of the Hamming distance of 32 vector elements as one MAC.
Moreover, we make no distinction between random and learned projection in the binarized \mimann, as computations remain the same.
The generation of the random projection on the device is not dominated by MAC computations and can be efficiently implemented with dedicated hardware accelerators~\cite{Kietzmann2020AIoT,Yang2015EfficientGenerator.}.


The computation of the number of MACs in the feature-based approach is not straight-forward to compute, mostly due to the matrix logarithm involved in the Riemannian feature extraction. 
We estimate the number of MACs of a matrix logarithm based on the complexity of an optimized Householder transformation $(\mathcal{O}(8 n^3_c / 3))$~\cite{burden1997numerical} and the iterative QR decomposition using implicit Wilkinson shift $(\mathcal{O}(6 n^3_c))$~\cite{wilkinson2013linear}. 

Table~\ref{tab:comp} shows that the computation of the features, in particular the computation of the covariance matrix, dominates the number of MACs in the Riemannian+SVM approach; the linear SVM makes up a negligible part (0.25\%) of the overall computations.
Conversely, the linear SVM occupies most of the memory footprint for storing the model parameters (80.38\%). 
As already stated in the memory footprint analysis, the introduction of the sparse bipolar random projection and binarized SVM reduces the memory footprint of the model; however, the overall number of MACs increases by 7.14$\times$.  
This yields a trade-off between lower complexity in inference (Riemannian+SVM) and lower memory footprint of the model (Riemannian+SVM binarized).

Among the considered CNN-based classifiers, TPCT requires 1.73\,GMACs per inference, which is more than one order of magnitude higher than all other classifiers. 
On the other side, EEGNet shows the lowest complexity with 13.14\,MMACs per inference. 
Compared to EEGNet, \mimann increases the total number of MACs only by 0.257\% in $d$=128 and by 0.523\% in $d$=256. 
The reason is that computations in EEGNet are dominated by the temporal convolution, making up 96\% of the computations. 
Consequently, replacing the fully connected classification layer by a projection layer of dimension $d$ has a negligible impact on the total number of MACs.
%
%


\begin{table}[]
\caption{Average classification accuracy, multiply-accumulate (MAC) operations per inference, and memory footprint of model weights in float16. 
}
\label{tab:comp}
\centering
\resizebox{\linewidth}{!}{%
\begin{threeparttable}
\begin{tabular}{lrrr}
\toprule
\textbf{Architecture}                & \textbf{Accuracy {[}\%{]}}                             & \multicolumn{1}{l}{\textbf{MAC/inf.}} & \multicolumn{1}{l}{\textbf{Mem. foot. {[}kB{]}}} \\ \hline
\rowcolor[HTML]{FFFFFF} 

\rowcolor[HTML]{FFFFFF} 
\rebuttal{\textbf{Riemannian+SVM}} & \rebuttal{\textbf{74.82}} & \rebuttal{\textbf{17.71\,M}}            & \rebuttal{\textbf{108.28}}                       \\
\rowcolor[HTML]{FFFFFF} 
\rebuttal{\textit{Bandpass filter}}                                &                 & \rebuttal{\textit{4\,138\,750}}             & \rebuttal{\textit{0.43}}                                \\
\rowcolor[HTML]{FFFFFF} 
\rebuttal{\textit{Covariance}}                             &                    & \rebuttal{\textit{9\,519\,125}}                  & \rebuttal{-}                     \\
\rowcolor[HTML]{FFFFFF} 
\rebuttal{\textit{Whitening}}                        &                     & \rebuttal{\textit{41\,624}}                      & \rebuttal{\textit{20.81}}                                \\
\rowcolor[HTML]{FFFFFF} 
\rebuttal{\textit{Matrix logarithm}}                        &                     & \rebuttal{\textit{3\,968\,155}$^\ddag$}                      & \rebuttal{-}                         \\
\rowcolor[HTML]{FFFFFF} 
\rebuttal{\textit{Linear SVM}}                        &                     & \rebuttal{\textit{43\,516}}                      & \rebuttal{\textit{87.04}}                                \\
\rowcolor[HTML]{dbdbdb} 
\rebuttal{\textbf{Riemannian+SVM binarized}} & \rebuttal{\textbf{73.55}} & \rebuttal{\textbf{126.47\,M}}            & \rebuttal{\textbf{71.24}}                       \\
\rowcolor[HTML]{dbdbdb} 
\rebuttal{\textit{Feature extraction}}                                &                 & \rebuttal{\textit{17\,667\,654}}             & \rebuttal{\textit{21.24}}                                \\
\rowcolor[HTML]{dbdbdb} 
\rebuttal{\textit{Projection}}                             &                    & \rebuttal{\textit{108\,790\,000}}                  &  \rebuttal{\textit{0.004$^*$}}         \\
\rowcolor[HTML]{dbdbdb} 
\rebuttal{\textit{Classification}}                             &                    & \rebuttal{\textit{12\,500}}                  &   \rebuttal{\textit{50.00}}                   \\
\rowcolor[HTML]{FFFFFF} 
\textbf{TPCT~\cite{Li2020AElectrodes}}          &\textbf{88.87}                     & \textbf{1.73\,G}                  & \textbf{15\,560}                               \\
\rowcolor[HTML]{dbdbdb}  
\textbf{CNN++~\cite{Zhao2017OnNetworks}}        & \textbf{81.10}                    & \textbf{18.24\,M}                  & \textbf{441.36}                                \\
\textbf{Shallow ConvNet~\cite{Schirrmeister2017DeepVisualization}} & \textbf{73.59}  & \textbf{62.99\,M}                 & \textbf{94.65}                               \\
\rowcolor[HTML]{dbdbdb}  
\textbf{EEGNet~\cite{Lawhern2018EEGNet:Interfaces}}    &  \textbf{72.32}           & \textbf{13.14\,M}                 & \textbf{5.26}                                \\
\rowcolor[HTML]{dbdbdb} 
\textit{Temp. Convolution}                             &             & \textit{12\,672\,000}             & \textit{1.09}                                \\
\rowcolor[HTML]{dbdbdb} 
\textit{Depw. Convolution}                          &                & \textit{396\,000}                 & \textit{0.83}                                \\
\rowcolor[HTML]{dbdbdb} 
\textit{Sep. Convolution}                           &                & \textit{71\,680}                  & \textit{1.15}                                \\
\rowcolor[HTML]{dbdbdb} 
\textit{FC}                                          &               & \textit{1088}                     & \textit{2.18}                                \\
\rowcolor[HTML]{FFFFFF} 
\textbf{\begin{tabular}[c]{@{}l@{}}\mimann\\ RP (LP) d=128\end{tabular}} & \textbf{70.41 (75.40)} & \textbf{13.17\,M}            & \textbf{3.59 (73.22)}                       \\
\rowcolor[HTML]{FFFFFF} 
\textit{Controller}                                &                 & \textit{13\,139\,680}             & \textit{3.07}                                \\
\rowcolor[HTML]{FFFFFF} 
\textit{Projection RP (LP)}                             &                    & \textit{34\,816}                  & \textit{0.004$^*$ (69.63)}                       \\
\rowcolor[HTML]{FFFFFF} 
\textit{Classification}                        &                     & \textit{160}                      & \textit{0.51}                                \\
\rowcolor[HTML]{dbdbdb} 
\textbf{\begin{tabular}[c]{@{}l@{}}\mimann\\ RP (LP) d=256\end{tabular}} & \textbf{72.31 (76.21)} & \textbf{13.21\,M}            & \textbf{4.10 (143.36)}                         \\
\rowcolor[HTML]{dbdbdb} 
\textit{Controller}                            &                     & \textit{13\,139\,680}             & \textit{3.07}                                \\
\rowcolor[HTML]{dbdbdb} 
\textit{Projection RP (LP)}                           &                      & \textit{69\,632}                  & \textit{0.004$^*$ (129.3)}                      \\
\rowcolor[HTML]{dbdbdb} 
\textit{Classification}                      &                       & \textit{288}                      & \textit{1.02}   \\
\rowcolor[HTML]{FFFFFF}  
\textbf{Q-EEGNet~\cite{Schneider2020Q-EEGNet:Interfaces}}   & \textbf{70.80}        & \textbf{13.14\,M}                 & \textbf{2.55}$^\dagger$ \\
\bottomrule
\end{tabular}
\begin{tablenotes}\footnotesize
\item $^*$Random projection matrix is regenerated during operation using random number generator with 32-bit seed. 
\item $^\dagger$ Int8 weights. 
\item \rebuttal{$^\ddag$ Estimation based on complexity $\mathcal{O}\left(\frac{27 n^3_c n_b}{3} \right)$.}  
\end{tablenotes}
\end{threeparttable}
}
\end{table}

\section{Conclusion}\label{sec:conclusion}
In this paper, we propose to \emph{binarize} real-valued features in common feature-based and CNN-based MI-BCI classification approaches to reduce their memory footprint for storing model parameters. 
In both approaches, random projections are the key enabler for successful binarization while ensuring similar accuracy as the full precision model. 
Yet, random projections do not increase the memory footprint because the weights in the projection matrix can be regenerated (rematerialized) by a random function on the fly. 

First, we binarize multi-spectral Riemannian features with sparse bipolar random projection and classify them with binarized SVM readout.
Experimental results on 4-class MI dataset of BCI Competition IV-2a show that our method binarizes real-valued features in the same dimensionality with 7.42\% accuracy loss compared to SVM models in float16.
Further increasing the dimensionality in binary space improves the accuracy of the binary model, which results in 1.27\% lower accuracy but at a 1.31$\times$ smaller memory footprint.

Second, we propose \mimann, as the first MANN architecture for MI-BCI, which generates compact binary vectors using CNN-based feature extractor; it includes EEGNet, the bipolar random projection, and the binary key-value memory for classification. 
It achieves similar accuracy as EEGNet in float16 precision (72.31\% vs. 72.32\%), while requiring a similar number of MAC operations and having a 1.29$\times$ smaller memory footprint.
Moreover, the accuracy alleviates to 76.21\% by allowing the projection to be learned, but this also requires 27.28$\times$ higher memory footprint. 
The introduction of \mimann allows for a cheap model update/extension on the device at the edge without requiring backpropagation algorithms nor increasing the memory footprint significantly, thanks to the binary representation of the key memory.

\section*{Acknowledgment}
This project was supported in part by ETH Research Grant 09 18-2, and by EU’s H2020 under grant no. 780215.






\begin{thebibliography}{10}
\providecommand{\url}[1]{#1}
\csname url@samestyle\endcsname
\providecommand{\newblock}{\relax}
\providecommand{\bibinfo}[2]{#2}
\providecommand{\BIBentrySTDinterwordspacing}{\spaceskip=0pt\relax}
\providecommand{\BIBentryALTinterwordstretchfactor}{4}
\providecommand{\BIBentryALTinterwordspacing}{\spaceskip=\fontdimen2\font plus
\BIBentryALTinterwordstretchfactor\fontdimen3\font minus
  \fontdimen4\font\relax}
\providecommand{\BIBforeignlanguage}[2]{{%
\expandafter\ifx\csname l@#1\endcsname\relax
\typeout{** WARNING: IEEEtran.bst: No hyphenation pattern has been}%
\typeout{** loaded for the language `#1'. Using the pattern for}%
\typeout{** the default language instead.}%
\else
\language=\csname l@#1\endcsname
\fi
#2}}
\providecommand{\BIBdecl}{\relax}
\BIBdecl

\bibitem{Ramadan2017BrainReview}
\BIBentryALTinterwordspacing
R.~A. Ramadan and A.~V. Vasilakos, ``{Brain computer interface: control signals
  review},'' \emph{Neurocomputing}, vol. 223, pp. 26--44, 2017.
\BIBentrySTDinterwordspacing

\bibitem{Saeedi2016AdaptiveReliability}
S.~Saeedi, R.~Chavarriaga, R.~Leeb, and J.~D.~R. Millan, ``{Adaptive Assistance
  for Brain-Computer Interfaces by Online Prediction of Command Reliability},''
  \emph{IEEE Computational Intelligence Magazine}, vol.~11, no.~1, pp. 32--39,
  2016.

\bibitem{Perdikis2018TheUsers}
\BIBentryALTinterwordspacing
S.~Perdikis, L.~Tonin, S.~Saeedi, C.~Schneider, and J.~d.~R. Mill{\'{a}}n,
  ``{The Cybathlon BCI race: Successful longitudinal mutual learning with two
  tetraplegic users},'' \emph{PLOS Biology}, vol.~16, no.~5, p. e2003787, 2018.
\BIBentrySTDinterwordspacing

\bibitem{Xiong2019AActivation}
M.~Xiong, A.~Brandenberger, M.~Bulger, W.~Chien, A.~Doyle, W.~Hao, J.~Jiang,
  K.~Kim, S.~Lahlou, C.~Leung \emph{et~al.}, ``{A low-cost, semi-autonomous
  wheelchair controlled by motor imagery and jaw muscle activation},'' in
  \emph{2019 IEEE International Conference on Systems, Man and Cybernetics
  (SMC)}.\hskip 1em plus 0.5em minus 0.4em\relax IEEE, 2019, pp. 2180--2185.

\bibitem{Cho2019MotorMethod}
W.~Cho, A.~Heilinger, R.~Ortner, J.~Swift, G.~Edlinger, C.~Guger, N.~Murovec,
  R.~Xu, M.~Zehetner, and S.~Schobesberger, ``{Motor Rehabilitation for
  Hemiparetic Stroke Patients Using a Brain-Computer Interface Method},'' in
  \emph{2018 IEEE International Conference on Systems, Man, and Cybernetics,
  (SMC)}.\hskip 1em plus 0.5em minus 0.4em\relax IEEE, 2019, pp. 1001--1005.

\bibitem{Tangermann2012ReviewIV.}
\BIBentryALTinterwordspacing
M.~Tangermann, K.-R. M{\"{u}}ller, A.~Aertsen, N.~Birbaumer, C.~Braun,
  C.~Brunner, R.~Leeb, C.~Mehring, K.~J. Miller, G.~R. M{\"{u}}ller-Putz
  \emph{et~al.}, ``{Review of the BCI Competition IV.}'' \emph{Frontiers in
  neuroscience}, vol.~6, p.~55, 2012.
\BIBentrySTDinterwordspacing

\bibitem{Lotte2018AUpdate}
\BIBentryALTinterwordspacing
F.~Lotte, L.~Bougrain, A.~Cichocki, M.~Clerc, M.~Congedo, A.~Rakotomamonjy, and
  F.~Yger, ``{A review of classification algorithms for EEG-based
  brain–computer interfaces: a 10 year update},'' \emph{Journal of Neural
  Engineering}, vol.~15, no.~3, p. 031005, 2018.
\BIBentrySTDinterwordspacing

\bibitem{Schirrmeister2017DeepVisualization}
\BIBentryALTinterwordspacing
R.~T. Schirrmeister, J.~T. Springenberg, L.~D.~J. Fiederer, M.~Glasstetter,
  K.~Eggensperger, M.~Tangermann, F.~Hutter, W.~Burgard, and T.~Ball, ``{Deep
  learning with convolutional neural networks for EEG decoding and
  visualization},'' \emph{Human Brain Mapping}, vol.~38, no.~11, pp.
  5391--5420, 2017.
\BIBentrySTDinterwordspacing

\bibitem{Hersche2018}
\BIBentryALTinterwordspacing
M.~Hersche, T.~Rellstab, P.~D. Schiavone, L.~Cavigelli, L.~Benini, and
  A.~Rahimi, ``{Fast and Accurate Multiclass Inference for MI-BCIs Using Large
  Multiscale Temporal and Spectral Features},'' in \emph{2018 26th European
  Signal Processing Conference (EUSIPCO)}.\hskip 1em plus 0.5em minus
  0.4em\relax IEEE, 2018, pp. 1690--1694.
\BIBentrySTDinterwordspacing

\bibitem{KaiKengAng2008FilterInterface}
\BIBentryALTinterwordspacing
{Kai Keng Ang}, {Zhang Yang Chin}, {Haihong Zhang}, and {Cuntai Guan},
  ``{Filter Bank Common Spatial Pattern (FBCSP) in Brain-Computer Interface},''
  in \emph{2008 IEEE International Joint Conference on Neural Networks (IEEE
  World Congress on Computational Intelligence)}.\hskip 1em plus 0.5em minus
  0.4em\relax IEEE, 2008, pp. 2390--2397.
\BIBentrySTDinterwordspacing

\bibitem{Barachant2013ClassificationApplications}
\BIBentryALTinterwordspacing
A.~Barachant, S.~Bonnet, M.~Congedo, and C.~Jutten, ``{Classification of
  covariance matrices using a Riemannian-based kernel for BCI applications},''
  \emph{Neurocomputing}, vol. 112, pp. 172--178, 2013.
\BIBentrySTDinterwordspacing

\bibitem{Brunner2008BCIA}
\BIBentryALTinterwordspacing
C.~Brunner, R.~Leeb, G.~R. M\"uller-Putz, A.~Schl\"ogl, and G.~Pfurtscheller,
  ``{BCI} competition 2008 - {G}raz data set {A},''
  \url{http://bnci-horizon-2020.eu/database/data-sets}.
\BIBentrySTDinterwordspacing

\bibitem{Li2020AElectrodes}
M.~A. Li, J.~F. Han, and L.~J. Duan, ``{A Novel MI-EEG Imaging with the
  Location Information of Electrodes},'' \emph{IEEE Access}, vol.~8, pp.
  3197--3211, 2020.

\bibitem{Lawhern2018EEGNet:Interfaces}
\BIBentryALTinterwordspacing
V.~J. Lawhern, A.~J. Solon, N.~R. Waytowich, S.~M. Gordon, C.~P. Hung, and
  B.~J. Lance, ``{EEGNet: a compact convolutional neural network for EEG-based
  brain–computer interfaces},'' \emph{Journal of Neural Engineering},
  vol.~15, no.~5, p. 056013, 2018.
\BIBentrySTDinterwordspacing

\bibitem{Rachkovskij2017BinaryEstimation}
\BIBentryALTinterwordspacing
D.~A. Rachkovskij, ``{Binary Vectors for Fast Distance and Similarity
  Estimation},'' \emph{Cybernetics and Systems Analysis}, vol.~53, no.~1, pp.
  138--156, 2017.
\BIBentrySTDinterwordspacing

\bibitem{Du2017ReservoirProcessing}
\BIBentryALTinterwordspacing
C.~Du, F.~Cai, M.~A. Zidan, W.~Ma, S.~H. Lee, and W.~D. Lu, ``{Reservoir
  computing using dynamic memristors for temporal information processing},''
  \emph{Nature Communications}, vol.~8, no.~1, p. 2204, 2017.
\BIBentrySTDinterwordspacing

\bibitem{Mountain2018MemristorProcessor}
D.~J. Mountain, M.~R. McLean, and C.~D. Krieger, ``{Memristor Crossbar Tiles in
  a Flexible, General Purpose Neural Processor},'' \emph{IEEE Journal on
  Emerging and Selected Topics in Circuits and Systems}, vol.~8, no.~1, pp.
  137--145, 2018.

\bibitem{Chakraborty2019Input-AwareDetection}
D.~Chakraborty, S.~Raj, S.~L. Fernandes, and S.~K. Jha, ``{Input-Aware
  Flow-Based Computing on Memristor Crossbars with Applications to Edge
  Detection},'' \emph{IEEE Journal on Emerging and Selected Topics in Circuits
  and Systems}, vol.~9, no.~3, pp. 580--591, 2019.

\bibitem{Burr2016RecentTechnology}
G.~W. Burr, M.~J. BrightSky, A.~Sebastian, H.~Y. Cheng, J.~Y. Wu, S.~Kim, N.~E.
  Sosa, N.~Papandreou, H.~L. Lung, H.~Pozidis \emph{et~al.}, ``{Recent Progress
  in Phase-Change Memory Technology},'' \emph{IEEE Journal on Emerging and
  Selected Topics in Circuits and Systems}, vol.~6, no.~2, pp. 146--162, 2016.

\bibitem{Karunaratne2020}
\BIBentryALTinterwordspacing
G.~Karunaratne, M.~Le~Gallo, G.~Cherubini, L.~Benini, A.~Rahimi, and
  A.~Sebastian, ``In-memory hyperdimensional computing,'' \emph{Nature
  Electronics}, vol.~3, no.~6, pp. 327--337, 2020.
\BIBentrySTDinterwordspacing

\bibitem{Saade2016RandomLight}
\BIBentryALTinterwordspacing
A.~Saade, F.~Caltagirone, I.~Carron, L.~Daudet, A.~Dremeau, S.~Gigan, and
  F.~Krzakala, ``{Random projections through multiple optical scattering:
  Approximating Kernels at the speed of light},'' in \emph{2016 IEEE
  International Conference on Acoustics, Speech and Signal Processing
  (ICASSP)}.\hskip 1em plus 0.5em minus 0.4em\relax IEEE,  2016, pp.
  6215--6219.
\BIBentrySTDinterwordspacing

\bibitem{Tan2016ApplyingBCI}
\BIBentryALTinterwordspacing
P.~Tan, W.~Sa, and L.~Yu, ``{Applying Extreme Learning Machine to
  classification of EEG BCI},'' in \emph{2016 IEEE International Conference on
  Cyber Technology in Automation, Control, and Intelligent Systems
  (CYBER)}.\hskip 1em plus 0.5em minus 0.4em\relax IEEE,  2016, pp. 228--232.
\BIBentrySTDinterwordspacing

\bibitem{Song2018SelectiveDecoding}
C.~Song, A.~Wang, F.~Lin, J.~Xiao, X.~Yao, and W.~Xu, ``{Selective CS: An
  Energy-Efficient Sensing Architecture for Wireless Implantable Neural
  Decoding},'' \emph{IEEE Journal on Emerging and Selected Topics in Circuits
  and Systems}, vol.~8, no.~2, pp. 201--210, 2018.

\bibitem{Kim2015Large-scaleHashing}
\BIBentryALTinterwordspacing
Y.~B. Kim and U.-M. O'Reilly, ``{Large-scale physiological waveform retrieval
  via locality-sensitive hashing},'' in \emph{2015 37th Annual International
  Conference of the IEEE Engineering in Medicine and Biology Society
  (EMBC)}.\hskip 1em plus 0.5em minus 0.4em\relax IEEE,  2015, pp. 5829--5833.
\BIBentrySTDinterwordspacing

\bibitem{Kim2016AnalysisSeries}
\BIBentryALTinterwordspacing
------, ``{Analysis of locality-sensitive hashing for fast critical event
  prediction on physiological time series},'' in \emph{2016 38th Annual
  International Conference of the IEEE Engineering in Medicine and Biology
  Society (EMBC)}.\hskip 1em plus 0.5em minus 0.4em\relax IEEE,  2016, pp.
  783--787.
\BIBentrySTDinterwordspacing

\bibitem{ZHANG2017BinaryMachine}
X.-m. Zhang, Y.-x. Dai, X.-b. Xu, and T.-t. He, ``{Binary Classification on
  ECoG Signals Using Optimized Extremely Learning Machine},'' \emph{DEStech
  Transactions on Computer Science and Engineering}, pp. 521--531, 2017.

\bibitem{Schneider2020Q-EEGNet:Interfaces}
\BIBentryALTinterwordspacing
T.~Schneider, X.~Wang, M.~Hersche, L.~Cavigelli, and L.~Benini, ``{Q-EEGNet: an
  Energy-Efficient 8-bit Quantized Parallel EEGNet Implementation for Edge
  Motor-Imagery Brain-Machine Interfaces},'' \emph{arXiv:2004.11690v1}, 2020.
\BIBentrySTDinterwordspacing

\bibitem{Hersche2020BinarySVM}
M.~Hersche, L.~Benini, and A.~Rahimi, ``{Binary Models for Motor-Imagery
  Brain-Computer Interfaces: Sparse Random Projection and Binarized SVM},''
  \emph{Proceedings - 2020 IEEE International Conference on Artificial
  Intelligence Circuits and Systems, AICAS 2020}, pp. 163--167, 2020.

\bibitem{Graves2014NeuralMachines}
\BIBentryALTinterwordspacing
A.~Graves, G.~Wayne, and I.~Danihelka, ``{Neural Turing Machines},''
  \emph{arXiv:1410.5401}, 2014.
\BIBentrySTDinterwordspacing

\bibitem{Laguna2019DesignNetworks}
A.~F. Laguna, M.~Niemier, and X.~S. Hu, ``{Design of Hardware-Friendly Memory
  Enhanced Neural Networks},'' \emph{Proceedings of the 2019 Design, Automation
  and Test in Europe Conference and Exhibition, DATE 2019}, pp. 1583--1586,
  2019.

\bibitem{Karunaratne2020_MANN}
\BIBentryALTinterwordspacing
G.~Karunaratne, M.~Schmuck, M.~L. Gallo, G.~Cherubini, L.~Benini, A.~Sebastian,
  and A.~Rahimi, ``{Robust High-dimensional Memory-augmented Neural
  Networks},'' \emph{arXiv:2010.01939}, pp. 1--32, 2020.
\BIBentrySTDinterwordspacing

\bibitem{Yger2017RiemannianReview}
\BIBentryALTinterwordspacing
F.~Yger, M.~Berar, and F.~Lotte, ``{Riemannian Approaches in Brain-Computer
  Interfaces: A Review},'' \emph{IEEE Transactions on Neural Systems and
  Rehabilitation Engineering}, vol.~25, no.~10, pp. 1753--1762, 2017.
\BIBentrySTDinterwordspacing

\bibitem{Zhao2017OnNetworks}
\BIBentryALTinterwordspacing
Y.~Zhao, S.~Yao, S.~Hu, S.~Chang, R.~Ganti, M.~Srivatsa, S.~Li, and
  T.~Abdelzaher, ``{On the improvement of classifying EEG recordings using
  neural networks},'' in \emph{2017 IEEE International Conference on Big Data
  (Big Data)}, 2017, pp. 1709--1711.
\BIBentrySTDinterwordspacing

\bibitem{Uran2019ApplyingAnalysis}
\BIBentryALTinterwordspacing
A.~Uran, C.~van Gemeren, R.~van Diepen, R.~Chavarriaga, and J.~d.~R.
  Mill{\'{a}}n, ``{Applying Transfer Learning To Deep Learned Models For EEG
  Analysis},'' \emph{arXiv:1907.01332}, 2019.
\BIBentrySTDinterwordspacing

\bibitem{Wang2020AnComputing}
\BIBentryALTinterwordspacing
X.~Wang, M.~Hersche, B.~T{\"{o}}mekce, B.~Kaya, M.~Magno, and L.~Benini, ``{An
  Accurate EEGNet-based Motor-Imagery Brain-Computer Interface for Low-Power
  Edge Computing},'' \emph{arXiv:2004.00077}, 2020.
\BIBentrySTDinterwordspacing

\bibitem{goldberger2000physiobank}
A.~L. Goldberger, L.~A.~N. Amaral, L.~Glass, J.~M. Hausdorff, P.~C. Ivanov,
  R.~G. Mark, J.~E. Mietus, G.~B. Moody, C.-K. Peng, and H.~E. Stanley,
  ``{PhysioBank, PhysioToolkit, and PhysioNet: components of a new research
  resource for complex physiologic signals},'' \emph{circulation}, vol. 101,
  no.~23, pp. e215--e220, 2000.

\bibitem{Pullini2018Mr.Processing}
\BIBentryALTinterwordspacing
A.~Pullini, D.~Rossi, I.~Loi, A.~Di~Mauro, and L.~Benini, ``{Mr. Wolf: A 1
  GFLOP/s Energy-Proportional Parallel Ultra Low Power SoC for IOT Edge
  Processing},'' in \emph{ESSCIRC 2018 - IEEE 44th European Solid State
  Circuits Conference (ESSCIRC)}.\hskip 1em plus 0.5em minus 0.4em\relax IEEE,
  2018, pp. 274--277.
\BIBentrySTDinterwordspacing

\bibitem{burden1997numerical}
R.~L. Burden and J.~D. Faires, ``Numerical analysis, brooks,'' \emph{Cole,
  Belmont, CA}, 1997.

\bibitem{wilkinson2013linear}
J.~H. Wilkinson, F.~L. Bauer, and C.~Reinsch, \emph{Linear algebra}.\hskip 1em
  plus 0.5em minus 0.4em\relax Springer, 2013, vol.~2.

\bibitem{Moakher2005AMatrices}
M.~Moakher, ``{A differential geometric approach to the geometric mean of
  symmetric positive-definite matrices},'' \emph{SIAM Journal on Matrix
  Analysis and Applications}, vol.~26, no.~3, pp. 735--747, 2005.

\bibitem{Bingham2001RandomReduction}
\BIBentryALTinterwordspacing
E.~Bingham and H.~Mannila, ``{Random projection in dimensionality reduction},''
  in \emph{Proceedings of the seventh ACM SIGKDD international conference on
  Knowledge discovery and data mining - KDD '01}.\hskip 1em plus 0.5em minus
  0.4em\relax New York, New York, USA: ACM Press, 2001, pp. 245--250.
\BIBentrySTDinterwordspacing

\bibitem{Johnson1984ExtensionsSpace}
W.~B. Johnson and J.~Lindenstrauss, ``{Extensions of Lipschitz mappings into a
  Hilbert space},'' \emph{Contemporary mathematics}, vol.~26, no.~1, pp.
  189--206, 1984.

\bibitem{Achlioptas2001Database-friendlyProjections}
\BIBentryALTinterwordspacing
D.~Achlioptas and {Dimitris}, ``{Database-friendly random projections},'' in
  \emph{Proceedings of the twentieth ACM SIGMOD-SIGACT-SIGART symposium on
  Principles of database systems - PODS '01}.\hskip 1em plus 0.5em minus
  0.4em\relax New York, New York, USA: ACM Press, 2001, pp. 274--281.
\BIBentrySTDinterwordspacing

\bibitem{Hersche2018ExploringInterfaces}
\BIBentryALTinterwordspacing
M.~Hersche, J.~d.~R. Mill{\'{a}}n, L.~Benini, and A.~Rahimi, ``{Exploring
  Embedding Methods in Binary Hyperdimensional Computing: A Case Study for
  Motor-Imagery based Brain-Computer Interfaces},'' \emph{arXiv:1812.05705},
  2018.
\BIBentrySTDinterwordspacing

\bibitem{McCloskey1989CatastrophicProblem}
\BIBentryALTinterwordspacing
M.~McCloskey and N.~J. Cohen, ``{Catastrophic Interference in Connectionist
  Networks: The Sequential Learning Problem},'' \emph{Psychology of Learning
  and Motivation}, vol.~24, pp. 109--165, 1989.
\BIBentrySTDinterwordspacing

\bibitem{Lake2015Human-levelInduction.}
\BIBentryALTinterwordspacing
B.~M. Lake, R.~Salakhutdinov, and J.~B. Tenenbaum, ``{Human-level concept
  learning through probabilistic program induction.}'' \emph{Science (New York,
  N.Y.)}, vol. 350, no. 6266, pp. 1332--8, 2015.
\BIBentrySTDinterwordspacing

\bibitem{Glorot2010UnderstandingNetworks}
X.~Glorot and Y.~Bengio, ``{Understanding the difficulty of training deep
  feedforward neural networks},'' \emph{Journal of Machine Learning Research},
  vol.~9, pp. 249--256, 2010.

\bibitem{Lotte2011RegularizingAlgorithms}
\BIBentryALTinterwordspacing
F.~Lotte and {Cuntai Guan}, ``{Regularizing Common Spatial Patterns to Improve
  BCI Designs: Unified Theory and New Algorithms},'' \emph{IEEE Transactions on
  Biomedical Engineering}, vol.~58, no.~2, pp. 355--362, 2011.
\BIBentrySTDinterwordspacing

\bibitem{Kietzmann2020AIoT}
\BIBentryALTinterwordspacing
P.~Kietzmann, T.~C. Schmidt, and M.~W{\"{a}}hlisch, ``{A Guideline on
  Pseudorandom Number Generation (PRNG) in the IoT},'' \emph{arXiv:2007.11839},
  pp. 1--23, 2020.
\BIBentrySTDinterwordspacing

\bibitem{Yang2015EfficientGenerator.}
\BIBentryALTinterwordspacing
G.~Yang, M.~D. Aagaard, and G.~Gong, ``{Efficient Hardware Implementations of
  the Warbler Pseudorandom Number Generator.}'' \emph{IACR Cryptology ePrint
  Archive}, vol. 2015, p. 789, 2015.
\BIBentrySTDinterwordspacing

\end{thebibliography}


\begin{IEEEbiography}[{\includegraphics[width=1in,height=1.25in,clip]{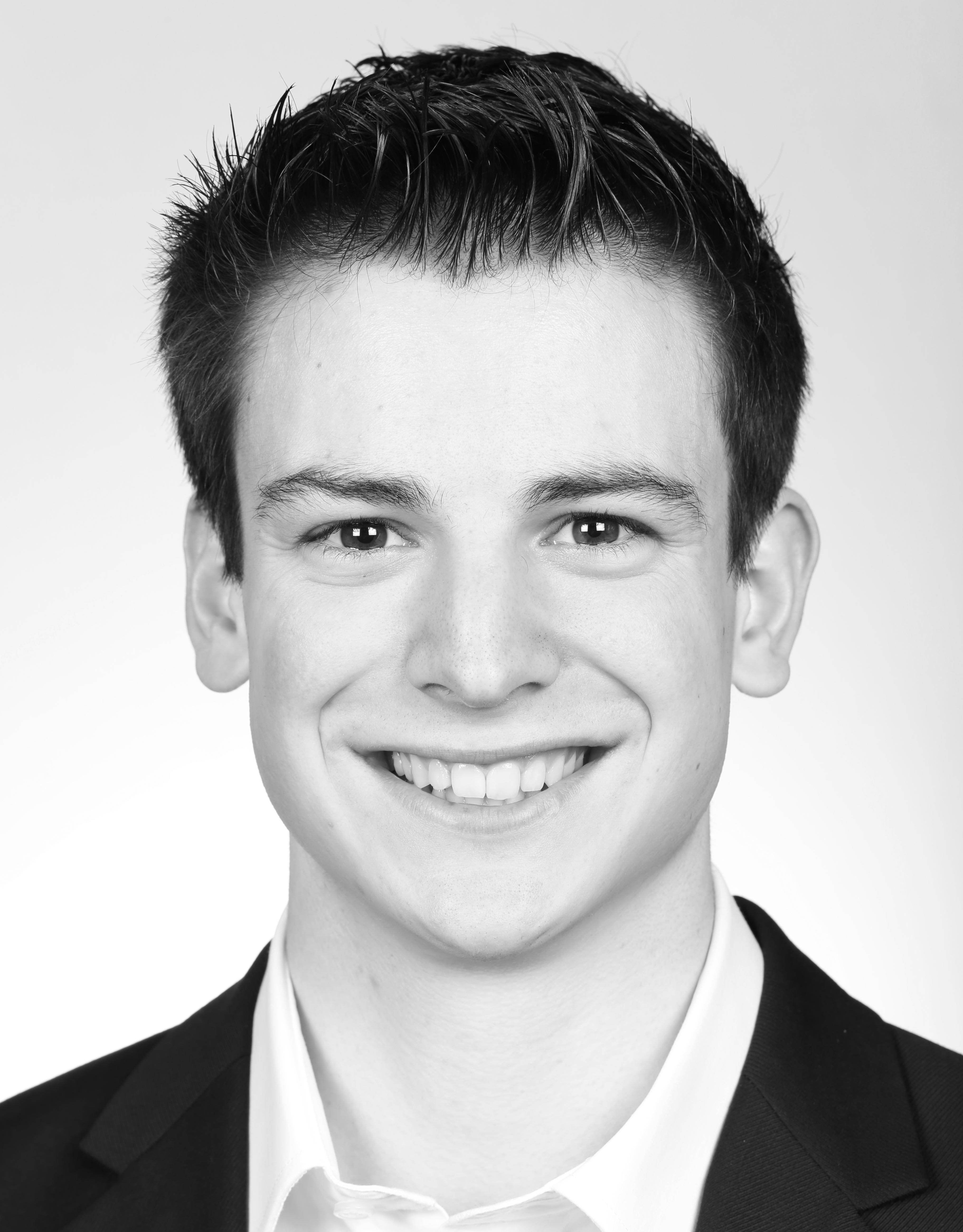}}]{Michael Hersche}
 received his M.Sc. degree from the Swiss Federal Institute of Technology Zurich (ETHZ), Switzerland, where he is currently pursuing a Ph.D. degree. 
Since 2019, he has been a research assistant at ETHZ in the group of Prof. Luca Benini at the Integrated Systems Laboratory. 
His research targets digital signal processing, artificial intelligence, and communication with focus on hyperdimensional computing.
Mr. Hersche received the 2020 IBM PhD Fellowship award. 
 \end{IEEEbiography}

 \begin{IEEEbiography}[{\includegraphics[width=1in,height=1.25in,clip]{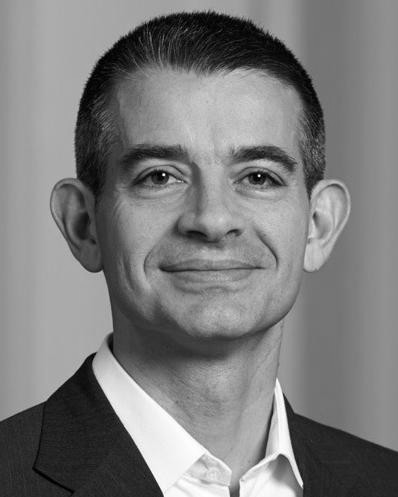}}]{Luca Benini}
has served as the Chief Architect for the Platform2012 at STMicroelectronics, Grenoble. 
He is currently the Chair of the Digital Circuits and Systems, ETH Zürich, and a Full Professor with the University of Bologna. 
His research interests are in the energy-efficient system and multi-core SoC design. 
He is also active in the area of energy-efficient smart sensors and sensor networks. 
He has published over 1000 articles in peer-reviewed international journals and conferences, four books, and several book chapters. 
He is a fellow of the ACM and a member of Academia Europaea.
 \end{IEEEbiography}
 
  \begin{IEEEbiography}[{
  \includegraphics[width=1in,height=1.25in,clip]{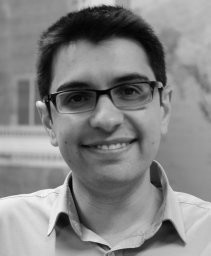}}]{Abbas Rahimi}
received the B.S. degree in computer engineering from the University of Tehran, Tehran, Iran, in 2010, and the M.S. and Ph.D. degrees in computer science and engineering from the University of California San Diego, La Jolla, CA, USA, in 2015, followed by postdoctoral researches at the University of California Berkeley, Berkeley, CA, USA, and at the ETH Zurich, Zurich, Switzerland. He is currently a Research Staff Member at the IBM Research-Zurich laboratory in Rüschlikon, Switzerland.
His research interests include brain-inspired hyperdimensional computing, neuro-symbolic AI, distributed embedded intelligent systems, and in general approximation opportunities in computation, communication, sensing, and storage with an emphasis on improving energy efficiency and robustness. Dr. Rahimi has received the ETH Zurich Postdoctoral Fellowship, and the 2015 Outstanding Dissertation Award in the area of ``New Directions in Embedded System Design and Embedded Software'' from the European Design and Automation Association. He was a co-recipient of the Best Paper Nominations at DAC (2013) and DATE (2019), and the Best Paper Awards at BICT (2017) and BioCAS (2018).
  \end{IEEEbiography}







\end{document}